\documentclass[
 reprint,
 amsmath,amssymb,
 aps, pra, floatfix, showkeys, nofootinbib, 
]{revtex4-2}
\usepackage[colorlinks]{hyperref}
\hypersetup{
    colorlinks,
    linkcolor={blue},
    citecolor={blue},
    urlcolor={blue}
}
\usepackage{graphicx}
\usepackage{dcolumn}
\usepackage{bm}
\usepackage{xcolor}
\usepackage{float}

\begin{document}
\preprint{APS/123-QED}

\title{Probability and fidelity of teleportation in a two-mode continuous-variable cluster state via a finite-resolution measurement device}

\author{J. A. Mendoza-Fierro}
\email{julio.mendoza@alumno.buap.mx}

\author{L. M. Ar\' evalo Aguilar}
\author{M. M. M\' endez Otero}
\affiliation{Facultad de Ciencias F\' isico Matem\' aticas, Benem\' erita Universidad Aut\' onoma de Puebla, Puebla, M\' exico.}
\email{larevalo@fcfm.buap.mx}
\date{\today}

\begin{abstract}
Continuous-variable measurements can not select individual outputs as in the discrete case; instead, the possible results are determined with a finite resolution. Then, it is said that continuous-variable measurement devices are insufficiently selective. By utilizing this concept, we show that the probability and fidelity of teleportation in a two-mode continuous-variable cluster state can be handled by both the localization and width of the selectivity interval of the measurement apparatus. Furthermore, we identify a trade-off relationship between the probability and fidelity of teleportation, which depends on both the width of the selectivity interval and the level of squeezing achieved in the cluster. Besides, we provide the mathematical expression for the probability distribution associated with the likelihood of teleportation in the two-mode cluster, which is a fundamental solution of the heat equation. In addition, we show that the fidelity of teleportation in the two-mode cluster is the quotient between the squared solution of a non-homogeneous heat equation and the solution of the conventional heat equation. We extend our approach to a configuration involving successive clusters with intermediate corrections between each teleportation step. To exemplify our proposal, we consider the specific case of a squeezed-coherent state as the quantum state under teleportation.
\end{abstract}
\maketitle
\section{Introduction}
Quantum computation (QC)\cite{Nielsen2000} exploits quantum phenomena such as entanglement and the principle of superposition to perform calculations and algorithms that surpass classical computers significantly. The revolutionary results include optimization problems \cite{Wang2023, Ajagekar2020}, fast simulation of complex quantum systems \cite{Daley2022, Georgescu2014}, and speed-up in integers factorization \cite{Shor1999}; thus, QC promises to improve enormously all generation of incoming technologies which will lay the foundation of unprecedented advancement in human understanding of the nature \cite{Deutsch2020}. Therefore, it is imperative to increase the research in all branches that drive advancements in QC. 

Currently, multiple models \cite{Wang2021} exist for the realization of universal QC \cite{Van2007}; this paper, however, deals with the building block of continuous variable (CV) one-way QC \cite{Raussendorf2001, Browne2006}. The stratagem of this scheme involves the implementation of unitaries and algorithms via local projective measurements on a highly multipartite entangled system known as a cluster state. Remarkably, this model allows skipping the requirement of unitary evolution, which drives the reduction of the number of necessary quantum gates for the processing of quantum information, which is carried principally by both the preparation of the entangled state and the structure of the measurements \cite{Gross2007}. Notably, one-way QC has been proven to be fault-tolerant \cite{Raussendorf2006, Menicucci2014, Fujii2010, Gottesman2001}, providing a robust basis for error mitigation; besides, one-way QC offers an inherently adaptive framework since diverse entanglement structures and distinct measurements schemes give the capacity to carry out different quantum algorithms \cite{Walther2005}.

Alternative to the employment of qubits for cluster construction in QC, we have the option of the continuous-variable (CV) approach \cite{Lloyd1999}, which weaves the cluster using quantum states expanded in a continuous eigenbasis \cite{Zhang2006}. The cornerstone for using these multipartite systems in universal QC was proposed in \cite{Menicucci2006} utilizing the adaptation of the one-qubit teleportation circuit \cite{Zhou2000, Nielsen2006a} to the CV regime through an optical insight. In this context, perfect teleportation only happens in the ideal situation where the sources building the cluster have infinite squeezing; however, in the real aim, only finite squeezing can be implemented, driving unavoidably to a Gaussian noise affecting the propagation of the information through the cluster \cite{Menicucci2006}.

Some of the current challenges to implementing efficient large-scale QC with CV cluster states involve seeking ways to deal with Gaussian imperfections due to finite squeezing and achieving a determined noise threshold for efficient operation. In the actual context, there are proposals for sidestep or lessen the cumulative noise due to the finite squeezing; for example, schemes of error correction \cite{Su2018, Hao2021} and surface code \cite{Fukui2018, Noh2020, Larsen2020, Larsen2021, Bourassa2021, Tzitrin2021, Noh2022, Korolev2022, Fukui2023} (see also \cite{Du2023} and references therein), error decreasing models by using a weighted controlled-$Z$ operation \cite{Zinatullin2022} and the cubic phase gate \cite{Zinatullin2022}, and the research in the evaluation of errors coming from different entangling operations \cite{Zinatullin2021}. However, in the current literature, no exists yet formalism—to the best of our knowledge—that addresses the handling of Gaussian noise through the measurement mechanism itself in one-way QC with CV cluster states; this fact is especially relevant since one-way QC relies principally on the measurement mechanism.
A first step to this goal is provided in Ref. \cite{Gu2009}, where it is shown that the Gaussian noise affecting the teleportation in a two-mode CV cluster state is strongly dependent on the measurement outcome; that is, there will be some measurement values for which the effect of Gaussian distortions will be more severe, will other maintain a higher fidelity of the propagated quantum information. Then, this fact establishes an interesting question:  \textit{Is it possible to use the measurement mechanism as a means to select a range of outcomes that allows handling the Gaussian noise affecting the propagation of information through a two-mode CV cluster?}. This question holds significant importance, as it enables us to delve into an aspect that—to the best of our knowledge—has not been considered yet regarding CV measurements in the teleportation process inside a two-mode CV cluster. 

The premise to establish a link between the measurement and the Gaussian noise affecting the teleportation process in a two-mode cluster ranges from the fact that projective CV measurements can not be infinitely selective \cite{Cohen1977, Pati2000}; that is, it can not select isolated outputs as in the case of observables with discrete eigenspectrum. Instead, CV measurements necessarily resolve  measurement results within a finite range which represents the power of resolution of the measurement device \cite{Cohen1977}. This feature makes finite-resolution measurements unique in the CV scenario, as they introduce an intrinsic probabilistic structure that directly impacts the performance of the protocol under consideration. Even further, the conception of a projective measurement in the CV regime can be misinterpreted, yielding a wrong formulation of a continuous-valued projection operator as a simple external product such as $\left| x \right\rangle \left\langle x\right|$ for $x$ valued continuously, as is done explicitly in Ref. \cite{Weedbrook2012} in the context of the teleportation through a two-mode CV cluster. Then, besides the mathematical difficulties of this formulation, it does not account for the probabilistic nature arising in CV teleportation processes \cite{Kumar2022}, as required by the quantum measurement postulate \cite{Paris2012}.
 
Notably, previous studies have reported probability–fidelity trade-offs in CV quantum teleportation schemes in the context of non-Gaussian operations \cite{Kumar2023, Navarrete2012, Su2011, Anno2010, Anno2007, Op2000}, where the trade-off arises from the probabilistic (heralded) preparation of non-Gaussian entangled resource states. In those approaches, the teleportation process itself—typically based on the original Braunstein–Kimble scheme—remains deterministic once the non-Gaussian resource is available, with the improvement in fidelity being attributed to the properties of the resource state. In contrast, we identify a qualitatively different trade-off in the teleportation model proposed in Ref. \cite{Menicucci2006}, originally introduced in the context of one-way quantum computation. To the best of our knowledge, such a trade-off has not been previously reported. Specifically, our results reveal a direct trade-off between teleportation fidelity and probability arising from the finite resolution of the measurement device.

Then, in this work, we answer the previously formulated question, showing how a finite-resolution measurement allows us to handle the quality (probability and fidelity) of teleportation in a two-mode CV cluster state presented in Ref. \cite{Menicucci2006}. Besides, we provide the mathematical expression for the momentum probability distribution associated with the measurement carried on the cluster, which is a solution of the heat equation. In addition, we employ the formalism of an insufficiently selective measurement in a scheme of linear sequential teleportations; particularly, we derive exact mathematical expressions giving both the probability and fidelity of teleportation. It must be noted that the concatenation of two-mode cluster states in a linear sequence has been considered in the original proposal for universal quantum computation \cite{Menicucci2006}, but also in Ref. \cite{Gu2009} from the perspective of the Wigner function (see also Ref. \cite{Menicucci2010}); these papers show that the teleportation down the linear cluster provides a resultant Gaussian noise dependent on both the measurement results and the Gaussian envelopes of each step of the sequence; as a difference of these works, our proposed configuration considers between each step of the sequence the set of intermediate corrections applied in the case of an ideal cluster which relies on infinite squeezing \cite{Weedbrook2012}. To clarify our proposal, we will appeal to a squeezed-coherent state as the quantum mode to be teleported through the cluster state. Furthermore, we show that in both the single and the recursive scheme, insufficiently selective measurements necessarily carry a probability of getting or not teleportation as is expected from the quantum measurement formalism \cite{Jordan2024}. Hence, the research of this paper provides a first path for engineering schemes of linear two-mode cluster states that allow the handling of the inherent Gaussian through the measurement mechanism itself. 

Through this paper, we use shot-noise units ($\hbar=2$), and the structure is organized as follows: In Sec. \ref{Sec:2}, we review the CV quantum gates building a two-mode cluster state; besides, we delve into the concept of an insufficiently selective measurement device. In Sec. \ref{Sec:2.3}, by considering the formalism of an insufficiently selective measurement, we board the noisy teleportation process in a two-mode CV cluster state; besides, we obtain the expressions governing both the probability and the fidelity of teleportation; moreover, we explain how an insufficiently selective measurement allows handling these two parameters in the cluster. For clarification, we board a squeezed-coherent state as the system to be teleported. In Sec. \ref{Sec:3}, we treat the situation of sequential teleportations through a linear chain of two-mode cluster states, where we board again the quality of teleportation of a squeezed-coherent state from the perspective of the insufficiently selective measurement. We close the article with the conclusions in Sec. \ref{Sec:4}.
\section{Preliminaries} \label{Sec:2}
In this section we review the pertinent CV quantum gates building the architecture of teleportation of a two-mode CV cluster state as is presented in Ref. \cite{Menicucci2006} (see also \cite{Weedbrook2012}). Here, we do not provide an extensive review of the properties of these gates; for more details, see \cite{Pieter2010}.
\subsection{Important Continuous-Variable quantum gates} \label{sec:2.1}
The scope concerned with the current work is the CV quantum states—also referred to as quantum modes, for short—which are described through a continuous superposition of the position or momentum eigenbasis; that is, $\int dq~\psi(q) \left| q\right\rangle$ or $\int dp~\psi(p) \left| p \right\rangle$, where the basis set $\left\lbrace \left| q \right\rangle \right\rbrace$ and $\left\lbrace \left| p \right\rangle \right\rbrace$ are termed as the computational and the conjugate basis respectively. These states are in a relationship through a Fourier transform, that is,
\begin{equation}
\left|q \right\rangle =\frac{1}{2\sqrt{\pi}}\int dp~e^{-\frac{i}{2}qp}\left|p \right\rangle, \label{eq:1}
\end{equation}
\begin{equation}
\left|p \right\rangle =\frac{1}{2\sqrt{\pi}}\int dq~e^{\frac{i}{2}qp}\left|q \right\rangle .\label{eq:2}
\end{equation}
Now, we recover the generalization of the Pauli $\hat{\sigma}_{x}$ and $\hat{\sigma}_{z}$ operators, that is, the  Heisenberg-Weyl operators $\hat{X}(r)$ and $\hat{Z}(s)$, which are generators of translations in the phase space; they are defined as
\begin{equation}
\hat{X}(r)=e^{-i r \hat{p}/2},\label{eq:3}
\end{equation}
\begin{equation}
\hat{Z}(s)=e^{i s \hat{q}/2}.\label{eq:4}
\end{equation}
These operators belong to the Heisenberg-Weyl group; moreover, they act on the position and momentum quadrature eigenstates of Eqs. \eqref{eq:1} and \eqref{eq:2} according to
\begin{equation}
\hat{X}(r) \left|q\right\rangle =\left| q+r\right\rangle,~~~\hat{Z}(s) \left| q\right\rangle = e^{i s \hat{q}/2}\left|q\right\rangle, \label{eq:5}
\end{equation}
\begin{equation}
\hat{X}(r) \left|p\right\rangle =e^{-i r \hat{p}/2}\left| p\right\rangle,~~~\hat{Z}(s) \left| p\right\rangle = \left| p + s\right\rangle. \label{eq:6}
\end{equation}
In the Heisenberg picture, these gates act on the quadrature operators according to
\begin{equation}
\hat{X}(r)^{\dagger}\hat{q}\hat{X}(r)=\hat{q}+r,~~~\hat{X}(r)^{\dagger}\hat{p}\hat{X}(r)=\hat{p}, \label{eq:7}
\end{equation}
\begin{equation}
\hat{Z}(s)^{\dagger}\hat{p}\hat{Z}(s)=\hat{p}+s,~~~\hat{Z}(s)^{\dagger}\hat{q}\hat{Z}(s)=\hat{q}. \label{eq:8}
\end{equation}
The $\hat{X}(r)$ and $\hat{Z}(s)$ gates are related by $\hat{X}(r)\hat{Z}(s)=e^{-irs/2}\hat{Z}(s)\hat{X}(r)$. Moreover, in terms of the displacement operator $\hat{D}(\alpha)=\exp\left( \alpha \hat{a}^{\dagger}-\alpha^{\ast}\hat{a} \right)$, the $\hat{X}(r)$ and $\hat{Z}(s)$ operators are $\hat{X}(r)=\hat{D}(r/2)$ and $\hat{Z}(s)=\hat{D}(is/2)$ respectively; then, the $\hat{X}(r)$ operator gives displacements in the real axis of the phase space while the $\hat{Z}(s)$ operator brings translations on the imaginary axis. A possible implementation of the displacement operators with optical elements involves imping an optical mode with a strong bright-coherent state on a beam splitter with a small coefficient reflection \cite{Pieter2010}.

On the other hand, we recover the Fourier transform gate, which is the generalization  of the Hadamard gate for qubits; this is defined as
\begin{equation}
\hat{F}=\exp\left(\frac{i\pi}{4}\right) \exp\left(\frac{i \pi}{2}\hat{a}^{\dagger}\hat{a}\right)=\exp\left[\frac{i \pi}{8}\left(\hat{q}^2 + \hat{p}^2 \right) \right]. \label{eq:9}
\end{equation}
In the phase-space description, the Fourier gate represents a $\pi/2$ rotation from one quadrature to another, that is,
\begin{equation}
\hat{F}^{\dagger} \hat{q} \hat{F}=-\hat{p},~~~\hat{F}^{\dagger} \hat{p} \hat{F}=\hat{q}. \label{eq:10}
\end{equation}
The effect of the Fourier gate on the quadrature eigenstates is
\begin{equation}
\hat{F}\left| q \right\rangle = \left| p \right\rangle,~~~ \hat{F}^{\dagger}\left| q \right\rangle = \left| -p \right\rangle,
\end{equation}
\begin{equation}
\hat{F}\left| p \right\rangle = \left| -q \right\rangle,~~~ \hat{F}^{\dagger}\left| p \right\rangle = \left| q \right\rangle; \label{eq:11.1}
\end{equation}
besides, it acts on the $\hat{X}(r)$ and $\hat{Z}(s)$ gates according to 
\begin{equation}
\hat{F}^{\dagger}\hat{Z}(s) \hat{F}=\hat{X}(s),~~~\hat{F}\hat{X}(r) \hat{F}^{\dagger}=\hat{Z}(r).\label{eq:12}
\end{equation}
The Fourier gate represents a simple phase shift of the corresponding mode up to an overall phase factor \cite{Pieter2010}; therefore, it can be implemented optically through a delay on the mode by a lossless material with a linear refraction index.

On the other hand, we have the controlled operation $\hat{C}_{Z}$, which is the quantum gate responsible for weaving the entanglement between the modes of the cluster; it is defined as
\begin{equation}
\hat{C}_{Z}=e^{\frac{i}{2} \hat{q}_{c} \otimes \hat{q}_{t}},\label{eq:13}
\end{equation}
where the labels in the exponential stand for (control $\otimes$ target). The $\hat{C}_{Z}$ gate causes displacements in phase-space on the target mode by a quantity determined by the position eigenvalue of the control mode. 

In the Heisenberg frame the $\hat{C}_{Z}$ gate carries the following transformations
\begin{equation}
\hat{C}_{Z}^{\dagger} \hat{q}_{c}\hat{C}_{Z}=\hat{q}_{c},~~~ \hat{C}_{Z}^{\dagger} \hat{q}_{t}\hat{C}_{Z}=\hat{q}_{t}, \label{eq:16}
\end{equation}
\begin{equation}
\hat{C}_{Z}^{\dagger} \hat{p}_{c}\hat{C}_{Z}=\hat{p}_{c}+\hat{q}_{t},~~~ \hat{C}_{Z}^{\dagger} \hat{p}_{t}\hat{C}_{Z}=\hat{p}_{t}+\hat{q}_{c}. \label{eq:17}
\end{equation}
Then, from Eqs.  \eqref{eq:16} and \eqref{eq:17}, we can observe that the control (target) quadrature defining the $\hat{C}_{Z}$ remains as constant in their dynamics while the corresponding conjugate quadrature is just displaced by the position quadrature of the target (control). This behavior is consistent with the type of quantum no demolition (QND) Hamiltonians \cite{Walls1994, Zhang2006, Furusawa2014} which have the form $\hat{H}_{Z}=-\mathcal{K} \left(\hat{q}_{i} \otimes \hat{q}_{j}\right)$ being $\mathcal{K}$  a coupling constant which it can experimentally adjust to one without generality loss. Then, the $\hat{C}_{Z}$ gate act just unitary entangling interaction \cite{Pfister2019}. The experimental implementation comes through several mechanisms. For instance, it can be done optically \cite{Ukai2014}, through beam splitter and inline squeezers \cite{Braunstein2005I, Yurke1985}, or with atomic ensembles \cite{Wang2011}. It is worth noting that another QND entangling interaction for cluster state generation can be accomplished through a SUM gate \cite{Yoshikawa2008}.
\subsection{Insufficiently selective measurement apparatus} \label{sec:2.2}
In this section, we delve into the concept of an insufficiently selective measurement device as is explained in \cite{Cohen1977} (Chapter 3, Sec. E2b). This concept is helpful for the correct mathematical definition of projective quantum measurements in the context of CV quantum systems. The ideas reviewed will find application in the design and analysis of noise affecting the scheme of teleportation utilizing two-mode CV cluster states.

\begin{figure}
\includegraphics[width=0.50\textwidth]{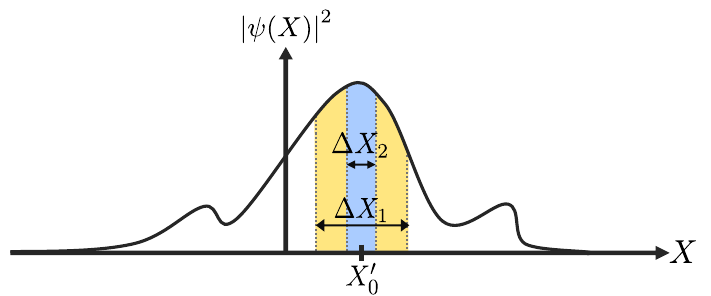}
\centering
\caption{Let be $\left|\psi(X)\right|^2=\left|\left\langle X\right.\left|\psi\right\rangle\right|^2$ the probability distribution (black curve with three peaks) of, for example, a continuous-variable observable $\hat{X}$. We consider a couple of detectors measuring $\hat{X}$ (such that $\hat{X}\left|X\right\rangle=X\left|X\right\rangle$) on the quantum system $\left|\psi\right\rangle$. These devices are centered on the point $X_{0}'$ on the $X$-axis, and can measure with resolutions $\Delta X_{1}$ and $\Delta X_{2}$; then, they can only detect points inside these intervals, giving indistinguishably the central value $X_{0}'$ as the measurement result. Besides, we have $\left|\Delta X_{1}\right| > \left|\Delta X_{2}\right|$. Then, while both detectors will give the outcome $X_{0}'$ when detecting points inside $\Delta X_{(1,2)}$, is the second which will provide a more accurate determination of the measured observable since all the set of points contained in $\Delta X_{2}$ are closer to $X_{0}'$ than the whole set contained in $\Delta X_{1}$. Notably, the probability of detecting $X_{0}'$ (represented by the blue and yellow areas below the curve and inside $\Delta X_{(1,2)}$) decreases as the measurement device becomes more accurate.}\label{fig:0}
\end{figure}

Let us suppose a measurement apparatus that provides binary outcomes such as ``yes'' or ``no'' (click or no click, 1 or 0, positive or negative, etc.). Let us assume that this apparatus can measure a physical quantity represented by an observable $\hat{A}$ of an arbitrary quantum state $\left|\psi\right\rangle$. If the measurement result falls inside an interval $\Delta$ on the real axis, the apparatus registers the ``yes" response. Conversely, if the measured value lies outside this interval, the recorded result is ``no." Then, such apparatus can determine measurement results within a range $\Delta$; in the following, this quantity will be termed as \textit{the power of resolution} or \textit{selectivity interval} of the measurement device. On the other hand, if the measurement device can detect individual outputs inside $\Delta$—that is, single eigenvalues of the observable $\hat{A}$—, we have a \textit{completely selective measurement apparatus}. Nevertheless, if $\Delta$ contains several eigenvalues of $\hat{A}$, the device lacks the resolution needed to distinguish each possible result; then, we say it is \textit{insufficiently selective}. Now, let us consider the selectivity interval centered around a value $X_{0}'$ for the insufficiently selective measurement case; then, all outcomes inside $\Delta$ will yield consistently $X_{0}'$, as they cannot be distinguished by the device. Therefore, the more selective the device is—i.e., the smaller the width of $\Delta$—the more accurate the determination of $X_{0}'$ becomes; see Fig. \ref{fig:0}.  For observables with continuous eigenspectrum—like the position or momentum of a quantum particle—\textit{it is not possible to have fully selective measurement devices giving precise measurement results} \cite{Cohen1977}\cite{Neumann1955}. It is important to note that this argument holds for any width of $\Delta$ since it will always contain an infinite number of eigenvalues, whatever their length on the real axis. Then, measurement apparatuses of observables with continuous eigenspectrum are always insufficiently selective \cite{Cohen1977}.

Let us return to the description of the measurement device as provided earlier; then, we will focus on projective measurements of an observable $\hat{X}$ with a continuous eigenspectrum. Be $\mathcal{S}_{\Delta X}$ the subspace of eigenstates of $\hat{X}$ whose eigenvalues fall in a determined interval $\Delta X$—that is, those measurement results for which the measurement device responses ``yes''—. Then, we have the projection operator $\hat{P}_{\Delta X}$, which projects the quantum state $\left|\psi\right\rangle$ on the subspace $\mathcal{S}_{\Delta X}$ around a central eigenvalue $X_{0}'$ inside the selectivity interval $\Delta X$ of the measurement device; mathematically, this operator is defined as
\begin{equation}
\hat{P}_{\Delta  X}=\int_{X_{0}' - \Delta X/2}^{X_{0}' +\Delta X/2} dX~\left|X\right\rangle \left\langle X \right|, \label{eq:18}
\end{equation}
where $\left|X\right\rangle$ represent an eigenstate of the observable $\hat{X}$; that is, $\hat{X}\left|X\right\rangle=X\left|X\right\rangle$ with $X \in \Delta X$, and $\left| X\right\rangle \in\mathcal{S}_{\Delta X}$. Besides, since $\hat{X}$ represent an observable, then $\left\lbrace \left| X \right\rangle \right\rbrace$ constitutes a basis set in the subspace $\mathcal{S}_{\Delta X}$. Hence, our projective measurement apparatus will respond ``yes''  for eigenstates—or superpositions of them—of $\hat{P}_{\Delta X}$ with the eigenvalue of $1$ and ``no'' for eigenstates of $\hat{P}_{\Delta X}$ with eigenvalue $0$.  It is worth noting that this measurement mechanism is essentially equivalent to measuring an observable with an infinitely degenerate eigenspectrum since there exists an infinity of eigenvalues inside $\Delta X$ for which the measurement device will give the response ``yes''. 
It is important to note that the definition of Eq. \eqref{eq:18} satisfies the adequate mathematical conditions for a projection operator; that is, $(\hat{P}_{\Delta X})^2=\hat{P}_{\Delta X}$ and $\hat{P}_{\Delta X} \left| X\right\rangle =\left|X\right\rangle ~\forall~ \left| X\right\rangle \in \mathcal{S}_{\Delta X}$ (we have $\hat{P}_{\Delta X} \left|X'\right\rangle =0 ~\forall~ \left|X'\right\rangle \notin \mathcal{S}_{\Delta X}$) as a difference of take the single external product $\left|X \right\rangle \left\langle X\right|$ which does not satisfy the previous requirements for projective measurements \cite{Pati2000}. 

Then, formally, there exists a set of indistinguishable states for which the measurement apparatus will yield the ``yes'' response (or, equivalently, the $X_{0}'$ outcome), and the corresponding eigenvalues will be probabilistically distributed inside its power of resolution as states the postulate of a quantum measurement \cite{Paris2012}. In the example of Fig. \ref{fig:0}, the likelihood for the ``yes'' answer is determined by the area under the curve describing the probability distribution of the measured observable inside the detector's selectivity region; mathematically, such likelihood is
\begin{equation}
\mathcal{P}_{\text{yes}}= \left\langle \psi\right|\hat{P}_{\Delta X} \left|\psi\right\rangle, \label{eq:19} 
\end{equation}
where evidently we have $\mathcal{P}_{\text{no}} =1 - \mathcal{P}_{\text{yes}}$ and $\left|\psi \right\rangle$ represents the quantum system under measurement.

On the other hand, the postulate of reduction of the (pure) quantum state for this measurement remains as: \textit{If for the measurement of an observable $\hat{X}$ on an arbitrary quantum state $\left|\psi\right\rangle$, we obtain the measurement result $X_{0}'$—that is, the ``yes answer—from the measurement device, the state $\left|\psi'\right\rangle$ of the system immediately after the measurement is the normalized projection of $\left|\psi\right\rangle$ on the subspace $\mathcal{S}_{\Delta X}$}, that is, 
\begin{equation}
\left|\psi'\right\rangle=\frac{\hat{P}_{\Delta X} \left|\psi\right\rangle}{ \sqrt{\left\langle\psi\right| \hat{P}_{\Delta X} \left|\psi\right\rangle}}. \label{eq:20} 
\end{equation}
Restricting ourselves to the realm of light detection, many insufficiently selective measurement devices are prevailing. For instance, photomultiplier tube devices can select impinging light photons in a determined range of wavelength depending on a diversity of factors such as the photo-sensible surface, the operation mode, and the quantum efficiency of its photocathode \cite{Polyakov2013}. Also, we have electron counting devices and all families of semiconductor diodes detecting light quanta, including avalanche, photo, and PIN diodes whose detection range is dependent on various factors; for example, the bandgap of the valence and conduction band of the considered semiconductor \cite{Maricau2013}, and even the surface area of the device. 
\section{Teleportation with finite squeezing in a two-mode continuous-variable cluster state} \label{Sec:2.3}
\subsection{Noisy Gaussian teleportation process} \label{2.3.1}
\begin{figure}
\includegraphics[width=0.50\textwidth]{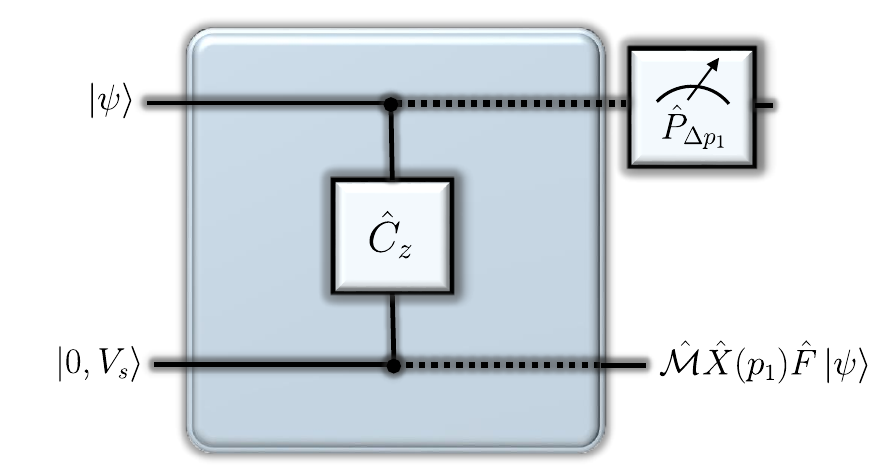}
\centering
\caption{Quantum circuit for noisy Gaussian teleportation of an arbitrary state $\left|\psi\right\rangle$ using a two-mode CV cluster state. The input states become entangled (dotted lines) through a $\hat{C}_Z$ gate. Subsequently, we perform an insufficiently selective projective measurement on the first mode; then, for all possible outputs inside the selectivity region of the measurement, we get the state $\mathcal{\hat{M}}\hat{X} (p_{1}) \hat{F} \left|\psi\right\rangle$ in the second mode of the circuit.}\label{fig:1}
\end{figure}

To elucidate the effect of a finite-resolution measurement on the teleportation process described in Ref. \cite{Menicucci2006}, we consider the quantum circuit of Fig. \ref{fig:1}. In the ideal scenario, we generate a two-mode CV cluster state through the entanglement of a zero-momentum eigenstate $\left|0\right\rangle_{p}$  with an arbitrary system $\left|\psi\right\rangle$ by utilizing a $\hat{C}_Z$ gate; then, a single projective measurement is performed on the mode containing the arbitrary state; through this, we can recover—through unitary corrections—the state $\left|\psi\right\rangle$ from the output of the mode that was initially associated with the zero-momentum eigenstate; see Ref. \cite{Weedbrook2012} for a detailed explanation. Crucially, within this framework, one can execute an algorithm $\hat{U}$ on the teleported state by simply performing measurements in the eigenbasis of the operator $\hat{U}^{\dagger} \hat{p} \hat{U}$, where $\hat{U}$ is a quantum operator that is diagonal in the computational basis and, consequently, commutes with the $C_Z$ gate. This process represents the basis for the measurement-based quantum computing model for continuous variables, allowing the concatenation of these kinds of circuits to build up a large-scale cluster state \cite{Weedbrook2012}. 
However, in practice, there is no physical way to generate zero momentum eigenstates since they constitute a non-normalizable set, which would need infinite energy to create them. Then, in substitution, these states are approximated by squeezed vacuum states, which—due to the finite squeezing—inevitably introduce a Gaussian noise that affects the teleportation of the state \cite{Menicucci2006}. In the following, we give a detailed analysis of this process by including the concept of an insufficiently selective measurement explained in Sec. \ref{sec:2.2}. 

We consider the circuit of Fig. \ref{fig:1}. Our input consists of an arbitrary state $\left|\psi\right\rangle$ on line 1 (top wire) and a squeezed vacuum state $\left|0, V_{s}\right\rangle$ on line 2 (bottom wire). We expand both states in position space according to
\begin{equation}
\left|\psi\right\rangle=\int dq~\psi(q) \left|q\right\rangle, \label{eq:21} 
\end{equation}
and
\begin{equation}
\left|0, V_{s}\right\rangle=\left(\frac{V_{s}^{2}}{2\pi} \right)^{\frac{1}{4}} \int dq~e^{-V_{s}^2 q^{2}/4} \left|q\right\rangle, \label{eq:22}
\end{equation}
where $V_{s}^{2}$ is a positive parameter related to the variances of the position and momentum probability distributions of the state according to $\delta_{q}^{2}=1/V_{s}^2$ and $\delta_{p}^{2}=V_{s}^2$ respectively; then, we can verify the saturation of the Heisenberg uncertainty relation $\delta_{q}^{2}\delta_{p}^{2}=1$; besides, if $\delta_{\hat{q}_{i}}^2$ ($\delta_{\hat{p}_{i}}^2$) $<1$, we have squeezing in the position (momentum) distribution. 
In an optical context, the primary method for the generation of the states of Eq. \eqref{eq:22} is the degenerate regime of the spontaneous parametric down-conversion \cite{Wu1986, Lvovsky2015}. Now, the initial state entering the circuit is the product $\left|\psi\right\rangle\left|0, V_{s}\right\rangle$; then, we apply the $\hat{C}_Z$ gate of Eq. \eqref{eq:13} to this state, obtaining the entangled state characterizing the cluster as
\begin{equation}
\begin{split}
 \hat{C}_{z}\left|\psi\right\rangle\left|0, V_{s}\right\rangle=&\left(\frac{V_{s}^{2}}{2\pi} \right)^{\frac{1}{4}} \int dq_{1} dq_{2}~\psi(q_{1})e^{-V_{s}^2 q_{2}^{2}/4}\\ &\times e^{\frac{i}{2} q_{1}q_{2}}\left|q_{1}\right\rangle \left|q_{2}\right\rangle.\label{eq:23}
 \end{split}
\end{equation}
\begin{figure}
\includegraphics[width=0.50\textwidth]{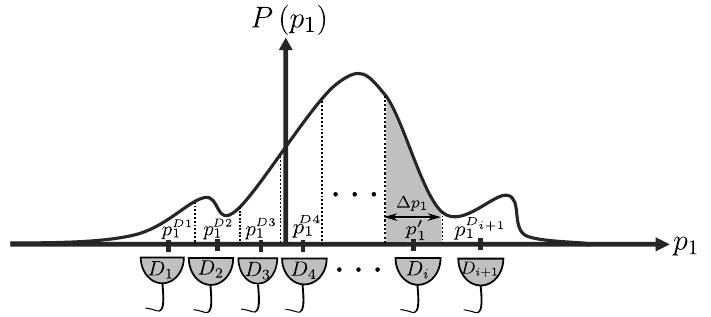}
\centering
\caption{We assume that the probability distribution of the measurement results within the cluster, $P(p_{1})$, is encompassed by a set of finite-resolution detectors, each having associated a measurement result, which is determined by the central coordinate of the corresponding selectivity interval. We identify the $i$-esim detector $D_{i}$ with an associated measurement result $p_{1}'$ and a selectivity interval $\Delta p_{1}$. From a phase-space perspective, each finite-resolution detector selects a region of the momentum marginal distribution $P(p_{1})$ (see Fig. \ref{fig:3} for a specific instance) associated with the quantum system under meaurement.
}\label{fig:00}
\end{figure}
The next step is to apply a momentum measurement on the first line of the cluster. In the original scheme of Ref. \cite{Menicucci2006} and in the Refs. \cite{Weedbrook2012, Gu2009}, the process is carried through a projective measurement operator of the form $\left|p_{1} \right\rangle\left\langle p_{1} \right|$, where $p_{1}$ is an arbitrary momentum eigenvalue of the first mode of the cluster state. This process implicitly assumes that the measurement performed can accurately select any momentum eigenvalue from the full eigenspectrum of measurement results. Instead, in our approach, we consider that the whole distribution of the measurement outputs is cover by a set of insufficiently selective measurement devices; see Fig. \ref{fig:00}. Then, appealing to the formalism presented in Section \ref{sec:2.2}, we will have a finite set of measurement results which will increase in both quantity and accuracy as the number of detectors increase. In particular, each detector  has its own detection range, given by its selectivity interval. The corresponding projective operator for the $i$-esim detector is of the following form
\begin{equation}
\hat{P}_{\Delta}^{(i)}=\hat{P}_{\Delta p_{1}}^{(i)} \otimes \mathbb{I}_{2},\label{eq:24}
\end{equation}
where $\mathbb{I}_{2}$ is the unity operator for the second mode, and $\hat{P}_{\Delta p_{1}}^{(i)}$ is given by 
\begin{equation}
\hat{P}_{\Delta p_{1}}^{(i)}= \int_{p_{1}' - \Delta p_{1}/2}^{{p_{1}' + \Delta p_{1}/2}} dp_{1} \left|p_{1}\right\rangle \left\langle p_{1}\right|, \label{eq:25}
\end{equation}
being $p_{1}'$ the central measurement result associated with the $i$-esim detector (see Section \ref{sec:2.2}). By applying $\hat{P}_{\Delta}^{(i)}$ to Eq. \eqref{eq:23}, and making use of $\left\langle p\right| \left. q \right\rangle=(2\sqrt{\pi})^{-1}e^{-iq p/2}$, we obtain
\begin{equation}
\begin{split}
&\hat{P}_{\Delta}^{(i)}\hat{C}_{z}\left|\psi\right\rangle \left|0, V_{s}\right\rangle=\left(\frac{V_{s}^{2}}{2\pi} \right)^{\frac{1}{4}} \left(2\sqrt{\pi} \right)^{-1}\int_{p_{1}' - \Delta p_{1}/2}^{{p_{1}' + \Delta p_{1}/2}} dp_{1}~  \\ 
&\times \int dq_{1} dq_{2}~\psi(q_{1})e^{-V_{s}^2 q_{2}^{2}/4} e^{\frac{i}{2}q_{1}q_{2}} e^{-\frac{i}{2}p_{1}q_{1}}\left|p_{1}\right\rangle\left|q_{2}\right\rangle, \label{eq:26}
\end{split}
\end{equation}
which can be rewritten as
\begin{equation}
\hat{P}_{\Delta }^{(i)}\hat{C}_{z}\left|\psi\right\rangle \left|0, V_{s}\right\rangle=\int_{p_{1}' - \Delta p_{1}/2}^{{p_{1}' + \Delta p_{1}/2}} dp_{1}~\left|p_{1}\right\rangle \left(\hat{\mathcal{M}}\hat{X}(p_{1}) \hat{F} \left|\psi\right\rangle \right),\label{eq:27}
\end{equation}
where
\begin{equation}
\hat{\mathcal{M}}= \int dq~f_{G}(q)\left|q\right\rangle \left\langle q\right|, \label{eq:29}
\end{equation}
is an operator which implies a Gaussian distortion.
According to Eq. \eqref{eq:20}, the normalized state after measurement is the normalized projection—i.e., divided by the constant $N=\left(\mathcal{P}_{\text{tel}}\right)^{1/2}$—of Eq. \eqref{eq:27} on the selectivity interval of the measurement. It is worth noting that the measurement mechanism utilized in our approach
corresponds to the measurement of an observable with a degenerate eigenspectrum \cite{Cohen1977}, since all points within the selectivity range of each detector correspond to its central value. Remarkably, under this conditions, the entanglement preserves on the selectivity region of the measurement (see, for example, Ref. \cite{Cohen2007} where measurements with degenerate spectrum can preserve entanglement); instead, a completely selective measurement unambiguously selects one momentum eigenstate, which drives that the second mode of the cluster unequivocally collapses to the single state: $N^{-1}\hat{\mathcal{M}}\hat{X}(p_{1}) \hat{F} \left|\psi \right\rangle$, and the entanglement is destroyed. 

After obtain the measurement result $p_{1}'$, certain information regarding the teleported state lives on the second mode of the cluster; we can obtain this state through the partial trace $\hat{\rho}_{2}=\text{Tr}_{1}\left[\hat{\rho}_{12} \right]$, where $\hat{\rho}_{12}=\hat{P}_{\Delta}\hat{C}_{z}\left|\psi\right\rangle\left|0, V_{s}\right\rangle \left\langle V_{s}, 0\right| \left\langle \psi \right|\hat{C}_{z}^{\dagger}\hat{P}_{\Delta}^{\dagger}$ is the global density operator of the system. Then, by utilizing  Eq. \eqref{eq:27}, and choosing a basis $\left\lbrace \left| p_{1}' \right\rangle \right\rbrace$ inside $\Delta p_{1}$, we obtain the normalized density operator of the second mode as
\begin{equation}
\hat{\rho}_{2}=N^{-2}\int_{p_{1}' - \Delta p_{1}/2}^{{p_{1}' + \Delta p_{1}/2}} dp_{1}~\hat{\mathcal{M}}\hat{X}(p_{1}) \hat{F} \hat{\rho}_{\text{in}} \hat{F}^{\dagger} \hat{X}^{\dagger}(p_{1})\hat{\mathcal{M}}^{\dagger},\label{eq:28}
\end{equation}
where $\hat{\rho}_{\text{in}}=\left|\psi\right\rangle \left\langle \psi \right|$ is the density operator of the state under teleportation, and $N$ is the normalization constant previously established. Notably, the use of finite squeezing in the teleportation process results inevitably in a Gaussian noise affecting the propagated information of the teleported quantum state \cite{Menicucci2006, Gu2009, Weedbrook2012}. This effect can be visualized by calculating the wave function in the second mode of the cluster for each momentum value inside the selectivity interval $\Delta p_{1}$; that is, by taking a position basis $\left\lbrace \left|q\right\rangle \right\rbrace$, we calculate $\psi'(q)=\left\langle  q \right|\hat{\mathcal{M}}\hat{X}(p_{1}) \hat{F}\left|\psi\right\rangle$, obtaining
\begin{equation}
\psi'(q)=\psi(q) f_{G}(q+p_{1}), ~~\forall ~ p_{1} \in \Delta p_{1}, \label{eq:30}
\end{equation}
where 
\begin{equation}
f_{G}(q)=\left( \frac{V_{s}^2}{2\pi}\right)^{\frac{1}{4}}e^{-V_{s}^2 q^2/4},\label{eq:31}
\end{equation}
is the wave function $f_{G}(q)=\left\langle q\right|\left.0,V_{s}\right\rangle$ coming from the squeezed vacuum state of Eq. \eqref{eq:23}.  Then, by considering an insufficiently selective measurement, the Eq. \eqref{eq:28} means that the post-measurement state in the second mode of the cluster carries the information of the teleported state modulated by a displaced Gaussian envelope defined on a range $\Delta p_{1}$ with center $p_{1}'$.

 On the other hand, the corresponding wave function in momentum space of Eq. \eqref{eq:30} can be obtained through a Fourier transform; that is,
\begin{eqnarray}
\begin{split}
\psi(p)&=N^{-1}\mathcal{F}\left\lbrace \psi(q)f_{G}(q+p_{1})\right\rbrace \\
&=N^{-1}\mathcal{F}\left\lbrace \psi(q)\right\rbrace \ast \mathcal{F}\left\lbrace f_{G}(q+p_{1})\right\rbrace, ~~\forall ~ p_{1} \in \Delta p_{1}; \label{eq:32}
\end{split}
\end{eqnarray}
therefore, $\psi(p)$ is given by the convolution between the wave functions in the momentum space of the teleported state and the squeezed vacuum state displaced by the measurement output.

Notably, since the measurement eigenspectrum associated with the first mode of the cluster is completely encompassed by a finite set of detectors, each device—corresponding to a specific measurement outcome—will have an associated detection probability. In the following subsection, we apply the formalism of an insufficiently selective measurement to derive the mathematical expression that describes the teleportation likelihood for each detector in our approach.
\subsection{Probability of teleportation} 
\label{2.3.2}
In our teleportation model within the two-mode CV cluster state, we assume that the whole measurement eigenspectrum is covered by a set of detectors with finite resolution; see Fig. \ref{fig:00}. If the $i$-esim detector registers a measurement, the outcome will be the central value $p_{1}'$ of their corresponding selectivity interval $\Delta p_{1}$. Then, since each device covers a portion of the total measurement eigenspectrum, each must have an associated probability of detection. In the following analysis, we derive the expression governing such likelihood.

The probability, $\mathcal{P}_{i}$, for obtaining a result inside the selectivity interval of the $i$-esim detector is given by Eq. \eqref{eq:19}; then, using this and Eq. \eqref{eq:27}, we obtain
\begin{equation}
\begin{split}
\mathcal{P}_{i} &=\left\langle \Psi\right| \hat{P}_{\Delta }^{(i)}  \left|\Psi\right\rangle   \\
&= \left\langle 0, V_{s}\right| \left\langle \psi\right| \hat{C}_{z}^{\dagger} \left(\hat{P}_{\Delta }^{(i)}\right)^{\dagger}  \hat{P}_{\Delta }^{(i)} \hat{C}_{z} \left|\psi\right\rangle\left|0, V_{s}\right\rangle\\
&= \int_{p_{1}' - \Delta p_{1}/2}^{{p_{1}' + \Delta p_{1}/2}} dp_{1} P(p_{1}),
\end{split}\label{eq:34}
\end{equation}
where in the second line we utilize $\left|\Psi \right\rangle=\hat{C}_{z}\left|\psi\right\rangle \left|0, V_{s}\right\rangle$ and the properties of the projection operator $\left(\hat{P}_{\Delta }^{(i)}\right)^{\dagger} =\hat{P}_{\Delta }^{(i)}$ and $\left(\hat{P}_{\Delta }^{(i)}\right)^{2}=\hat{P}_{\Delta }^{(i)}$. Besides,
\begin{equation}
\begin{split}
P(p_{1})&= \left\langle \psi \right| \hat{F}^{\dagger}\hat{X}^{\dagger}(p_{1})\hat{\mathcal{M}}^{\dagger}\hat{\mathcal{M}}\hat{X}(p_{1})\hat{F} \left| \psi\right\rangle \\
&=\int dq~\left[f_{G}(q)\right]^2 \rho(q-p_{1})
\end{split} \label{eq:34.1}
\end{equation}
represents the momentum probability distribution associated with the first mode of the cluster; moreover, $\rho(q)=\left|\psi(q) \right|^2$ is the probability distribution of the teleported state in position space, and $f_{G}(q)$ is given by Eq. \eqref{eq:31}. Hence, $P(p_{1})$ is determined from the convolution between the probability distributions of the squeezed vacuum state and that of the teleported state. Remarkably, by setting the variance  $\delta_{q}^2=2t$ for some $t>0$ and taking the explicit dependence of $f_{G}(q)$ on this parameter, we will have $\left[f_{G}(q,t)\right]^2=k(q,t)$, where $k(q,t)=(4\pi t)^{-1/2}\exp[-q^2/4t]$ represents the kernel for the generalized Weierstrass transform, which constitutes the fundamental solution of the heat equation\footnote{In fact, one can show through simple derivation that the integral of Eq. \eqref{eq:34.1} is also a solution of the heat equation.}\cite{Zayed1996}. Hence, the momentum probability distribution of the first mode of the cluster is given by the mirror image of the generalized Weierstrass transform of $\rho(q)$, which is simply a smoothed version of the position probability distribution of the teleported state. 

On the other hand, Eq. \eqref{eq:34} implies that the probability of teleportation in the cluster can be principally manipulated by the degree of squeezing of the squeezed vacuum state. To understand this argument, consider that in the limit $t\longrightarrow 0$ (that is, increasing the squeezing in position of the squeezed vacuum state), the kernel $k(q,t)$ represents a Dirac delta; consequently, the integral of Eq. \eqref{eq:34.1} will approximate the mirror image of the position probability distribution of the teleported state; thus, the integral of Eq. \eqref{eq:34} will determines the probability of teleportation for the detector. If the probability distribution $P(p_{1})$ lives entirely on $\Delta p_{1}$ we will have $\mathcal{P}_{\text{tel}}=1$; however, if the selectivity interval of the $i$-esim detector encompasses a fraction of $P(p_{1})$, then $\mathcal{P}_{\text{tel}}<1$. In real experimental scenarios, finite squeezing is the only available option; then, in this regime, the integral of Eq. \eqref{eq:34.1} represents a smoothed version of $\rho(q)$, which mandatory implies $\mathcal{P}_{\text{tel}}<1$, even in the situation where the measurement device completely encompasses the measurement eigenspectrum.

Now, in the following Subsection we analyze the implications of our approach in the quality (fidelity) of teleportation in a two-mode CV cluster.

\subsection{Fidelity of teleportation}\label{fidelof}
The objective of any teleportation process is to transmit the information of an unknown quantum system to a distant location without disturbance; however, achieving this in realistic terms is challenging due to degrading effects that affect the information propagation. A straightforward method  to quantify the quality of a teleportation process is by assessing its fidelity  $\mathcal{F}$; for a single pure state, this measure is defined as \cite{Braunstein2000}:
\begin{equation}
\mathcal{F}\equiv \left\langle \psi_{\text{in}} \right| \hat{\rho}_{\text{out}} \left| \psi_{\text{in}} \right\rangle,\label{eq:35} 
\end{equation} 
which is bounded according to $0\leq \mathcal{F} \leq 1$. The fidelity of teleportation of Eq. \eqref{eq:35} offers a direct mechanism for assessing the impact of finite squeezing and measuring the efficiency of information propagation within the cluster. On the other hand, in Ref. \cite{Gu2009}, it is shown that the effect of the Gaussian noise on the information propagation within the cluster is strongly dependent on the measurement results; that is, there will be some measurement outcomes for which the resulting distorted state will be closest to the original state, while others will provide less information about it.
Then, ranging from this idea, we focus on showing how a configuration of insufficiently selective measurement detectors can be harnessed to enhance the fidelity of teleportation in a two-mode CV cluster state.

To begin, we recall that the post-measurement density operator for the second mode of the cluster is given by Eq. \eqref{eq:28}. In the subsequent, we consider that the measurement eigenspectrum is encompassed by a set of high-resolution measurement devices, which can accurately resolve the momentum outcomes within their selectivity intervals\footnote{According to the formalism presented in Sec. \ref{sec:2.2}, this implies to take the selectivity interval of each detector small enough in order that the measurement result be closer the central value of the corresponding interval $\Delta p_{1}$.}. We call this the \textit{the quasi-selective approximation}. An instance of this approach can be found in the well-known balanced homodyne detection technique \cite{Lvovsky2009}, where the quadrature of an optical signal is determined by comparing it with a strong pump classical field called the local oscillator (LO). In this scenario, the two signals interfere on a beam splitter, and the output fields are measured by photodetectors, which register the incident photon flux; then, the difference between the resulting photocurrents provides information regarding the quadrature of the signal field for a determined phase of the LO. In this scheme, the photodetectors determine the number of light quanta within a spatiotemporal selectivity interval defined by their surface area and response time \cite{Smithey1993}. Therefore, the quasi-selective approximation is applicable when the photodetectors resolve—with high efficiency—a steady photon number for the measured signal field within their spatiotemporal selectivity interval. Besides, it is worth noting that the characteristics of the measured statistics also depend on the properties of the quantum state under measurement \cite{Raymer2004}. Then, by considering the quasi-selective approach, the integrands of Eqs. \eqref{eq:28} and \eqref{eq:34} becomes approximately constant on $\Delta p_{1}$ and the density operator of Eq. \eqref{eq:28} reduces to
\begin{equation}
\hat{\rho}_{2}\approx \left[P(p_{1})\right]^{-1} \hat{\mathcal{M}} \hat{X}(p_{1})\hat{F} \hat{\rho}_{\text{in}} \hat{F}^{\dagger} \hat{X}^{\dagger}(p_{1})\hat{\mathcal{M}}^{\dagger},\label{eq:36} 
\end{equation}
where $P(p_{1})$ is given by Eq. \eqref{eq:34.1}. Now, to obtain the output density operator, we follow the ideal scenario with infinite squeezing (see section B1 of \cite{Weedbrook2012}). In this scheme, the state in the second mode of the cluster collapses to $\left| \psi '\right\rangle = \hat{X}(p_{1})\hat{F}\left| \psi\right\rangle$; then, the teleportation process is complete \textit{until} we apply the corrections: $\hat{F}^{\dagger}\hat{X}^{\dagger}(p_{1})$ to $\left| \psi '\right\rangle$ (implying classical communication of the measurement output); hence, following this approach, we obtain the following output density operator
\begin{equation}
\hat{\rho}_{\text{out}}=\hat{F}^{\dagger}\hat{X}^{\dagger}(p_{1})\hat{\rho}_{2}\hat{X}(p_{1}) \hat{F}. \label{eq:37}
\end{equation}
This state carries an uncorrectable noise induced by the Gaussian distortion operator $\mathcal{\hat{M}}$ of Eq. \eqref{eq:29} which is not unitary; that is, we cannot take $\hat{\mathcal{M}}^{\dagger} \hat{\mathcal{M}} \neq \hat{\mathbb{I}}$. By substituting \eqref{eq:37} in Eq. \eqref{eq:35}, we obtain the fidelity of teleportation in the $i$-esim detector as
\begin{equation}
\mathcal{F}_{i}= \frac{1}{P(p_{1})}\left[\int dq~f_{G}(q) \rho(q-p_{1})\right]^2, \label{eq:38.2}
\end{equation}
where $f_{G}(q)$ is the Gaussian function of Eq. \eqref{eq:31}, $\rho(q)=\left|\psi(q)\right|^2$ is the position probability distribution of the state under teleportation, and $p_{1}$ represent an arbitrary measurement result. Referring the reader to Section \ref{2.3.2}, the Gaussian function $f_{G}(q)$ is related to the kernel of the generalized Weierstrass transform according to $f_{G}(q) = \left[k(q, t)\right]^{1/2}$. Then, instead of the usual heat equation, the integral in the numerator of Eq. \eqref{eq:38.2} is a solution of the following non-homogeneous version:
\begin{equation}
\frac{\partial^{2}u(p_{1}, t)}{\partial p_{1}^2}- \frac{\partial u(p_{1}, t)}{\partial t} =\left(\frac{q + p_{1}}{4t}\right) \frac{\partial u(p_{1}, t)}{\partial p_{1}}, \label{eq:39}
\end{equation}
with the Dirichlet boundary condition
\begin{equation}
u(p_{1},0)=f(-p_{1}). \label{eq:40}
\end{equation}
Therefore, the fidelity of teleportation associated with the $i$-esim detector (in the quasi-selective approach) is given by the quotient between the squared solution of the non-homogeneous heat equation of Eq. \eqref{eq:39} and the solution of the conventional heat equation \footnote{This argument even remains valid for any chosen measurement basis since any resulting diagonal operator can be absorbed in the probability distribution $\rho(q-p_{1})$ of the teleported state.}. 

On the other hand, the fidelity of Eq. \eqref{eq:38.2} refers to a single teleportation process; however, depending on the context, several teleportations may be necessary to extract the statistic of the quantum information processed by the cluster; then, in this situation, the average state teleported is mixed \cite{Gu2009}.  This state is obtained by averaging the output density operator of Eq. \eqref{eq:37} with the momentum probability distribution of Eq. \eqref{eq:34.1}; that is,
\begin{equation}
\hat{\rho}_{\text{out}}^{\text{av}}=\int_{-\infty}^{\infty} dp_{1}~P(p_{1}) \hat{\rho}_{\text{out}}, \label{eq:40.1}
\end{equation}
where the superscript `av' means `average'. By utilizing Eqs. \eqref{eq:35} and \eqref{eq:40.1}, the averaged fidelity on multiple teleportation processes is
\begin{equation}
\mathcal{F}^{\text{av}}=\int_{-\infty}^{\infty} dp_{1} ~\left[\int dq ~f_{G}(q) \rho(q-p_{1}) \right]^2.
 \label{eq:40.22}
\end{equation}
However, by focusing on the $i$-esim detector under the quasi-selective approach, and by making use of Eq. \eqref{eq:38.2}, we can express the average fidelity as
\begin{equation}
\mathcal{F}^{\text{av}}=P(p_{1})\Delta p_{1}\mathcal{F}_{i};\label{eq:40.23}
\end{equation}
then, the average fidelity becomes directly proportional to both the momentum probability distribution of the first mode of the cluster and the fidelity of individual teleportations. Since these quantities directly depend on the overlap between $f_{G}(q)$ and $\rho(q)$, it follows that the average fidelity is also determined by this overlap. Consequently, certain measurement outcomes will yield higher quality in teleportation, both in individual processes and on average. Given our consideration of Fig. \ref{fig:00}, we can design a filter-based experimental configuration that discards all measurements from those detectors whose associated outcomes do not provide higher-quality teleportation while allowing only the activation of that or those whose measurement results do. However, it is worth noting that for this mechanism, we must have previous knowledge of the state to teleport to select those devices whose associate measurement result offers the higher overlap between the functions $f_{G}(q)$ and $\rho(q)$. However, although any theoretical framework for teleportation demands an arbitrary quantum state under processing \cite{Bennet1993}, the experimental implementations commonly utilize well-characterized quantum states; for example, both satellite-based quantum communication \cite{Gonzalez2024} and quantum teleportation networks \cite{Shi2023} rely on the extensively researched Gaussian states \cite{Weedbrook2012}. Furthermore, large CV cluster states can be constructed by entangling  Gaussian states via beam splitters \cite{Zhang2006}. Then, although our proposed mechanism for enhancing the fidelity of teleportation requires knowledge of the quantum state being teleported, it retains its generality in the sense that it is applicable to \textit{any} quantum state under processing, via the selection of appropriate selectivity intervals tailored to the specific characteristics of the state in question. In the following Subsection, we illustrate this concept by applying it to a squeezed-coherent state as the system under processing.
\subsection{Teleportation of a squeezed-coherent state} \label{squeezed}
To get insight into the mechanism for enhancing the fidelity of teleportation via a set of insufficiently selective measurement devices, we center our analysis on the realm of individual teleportations; then, we must evaluate the Eqs. \eqref{eq:34} and \eqref{eq:38.2} for a specific input state. We consider a squeezed-coherent state $\left|\alpha, \xi \right>$, which is generated according to $\left|\alpha, \xi \right>=\hat{D}(\alpha) \hat{S}(\xi) \left| 0 \right>$, where $\hat{D}(\alpha)=\exp\left( \alpha \hat{a}^{\dagger} - \alpha^{\ast} \hat{a}\right) $ and $\hat{S}(\xi)=\exp[-\left(\xi (\hat{a}^{\dagger})^2 + \xi^{\ast} \hat{a}^2 \right)]$ are the displacement and the one-mode squeezing operators respectively; besides, we have the complex amplitude $\alpha=(q_{0} + i p_{0})/2$ and $\xi=re^{i \theta}/2$, being $q_{0}$ and $p_{0}$ the center of coordinates of the state under teleportation in the phase space and $r$ and $\theta$ the squeezing and rotation parameters for $r \in \mathbb{R}$ and $0\leq \theta \leq 2 \pi$. 

The wave function of a squeezed-coherent state has been determined by several methodologies \cite{Munguia2021}; at this step, we recover it ($\hbar=2$) in the position space:
\begin{equation}
\begin{split}
\psi_{\xi, \alpha}(q)=\frac{\left(1/2\pi \right)^{\frac{1}{4}}\sqrt{\cosh r_{1} - e^{-i\theta}\sinh r_{1}}}{\sqrt{\left|\cosh r_{1} - e^{i\theta}\sinh r_{1}\right|\left|\cosh r_{1} - e^{-i\theta}\sinh r_{1}\right|}} \\
\times e^{\frac{i}{2} p_{0}\left(q-\frac{q_{0}}{2}\right)}  \exp\left[-\frac{1}{4} \frac{\cosh r_{1} + e^{i\theta}\sinh r_{1}}{\cosh r_{1} - e^{i\theta}\sinh r_{1}} (q-q_{0})^{2} \right],
\end{split} \label{eq:41}
\end{equation}
where $r_{1} \in \mathbb{R}$ is its squeezing parameter. The variances of the quadratures of position and momentum of this state are
\begin{equation}
\delta_{q}^{2}(r_{1}, \theta)=\cosh 2 r_{1} - \cos \theta \sinh 2 r_{1},  \label{eq:42}
\end{equation}
\begin{equation}
\delta_{p}^{2}(r_{1}, \theta)= \cosh 2 r_{1} + \cos \theta \sinh 2 r_{1}.  \label{eq:43}
\end{equation}
Now, it must be noted that for the calculations of both the probability and the fidelity of teleportation (see Eqs. \eqref{eq:34.1} and \eqref{eq:38.2}), we need the position probability distribution of the  teleported state displaced by  $p_{1}$; then, we include such value together the central coordinate of position of the squeezed-coherent state in order to define an effective displacement  $X_{0}=q_{0} + p_{1}$, with $X_{0} \in \mathbb{R}$; hence, using the wave function of Eq. \eqref{eq:41} and considering the Eq. \eqref{eq:42}, we obtain the following displaced position probability distribution of the squeezed-coherent state
\begin{equation}
\rho(q-p_{1})=\left[2\pi \delta_{q}^{2}(r_{1}, \theta)\right]^{-\frac{1}{2}}e^{-\frac{(q-X_{0})^2}{2\delta_{q}^{2}(r_{1}, \theta)}}. \label{eq:44}
\end{equation}
In the following sections, we will utilize Eq. \eqref{eq:44} to analyze the teleportation process of the squeezed-coherent state through the two-mode CV cluster.
\subsubsection{Probability of teleportation} \label{2.4.1}
Let us calculate the measurement probability distribution associated with the first mode of the cluster by considering the teleportation of a squeezed-coherent state. By utilizing both the Gaussian function of Eq. \eqref{eq:31} and the probability distribution of Eq. \eqref{eq:44} in Eq. \eqref{eq:34.1}, we obtain
\begin{equation}
P(p_{1})=\left( \frac{\sigma^2}{2\pi }\right)^{\frac{1}{2}} e^{-\sigma^2 X_{0}^2/2}, \label{eq:44.01}
\end{equation}
\begin{figure}
\includegraphics[width=0.45\textwidth]{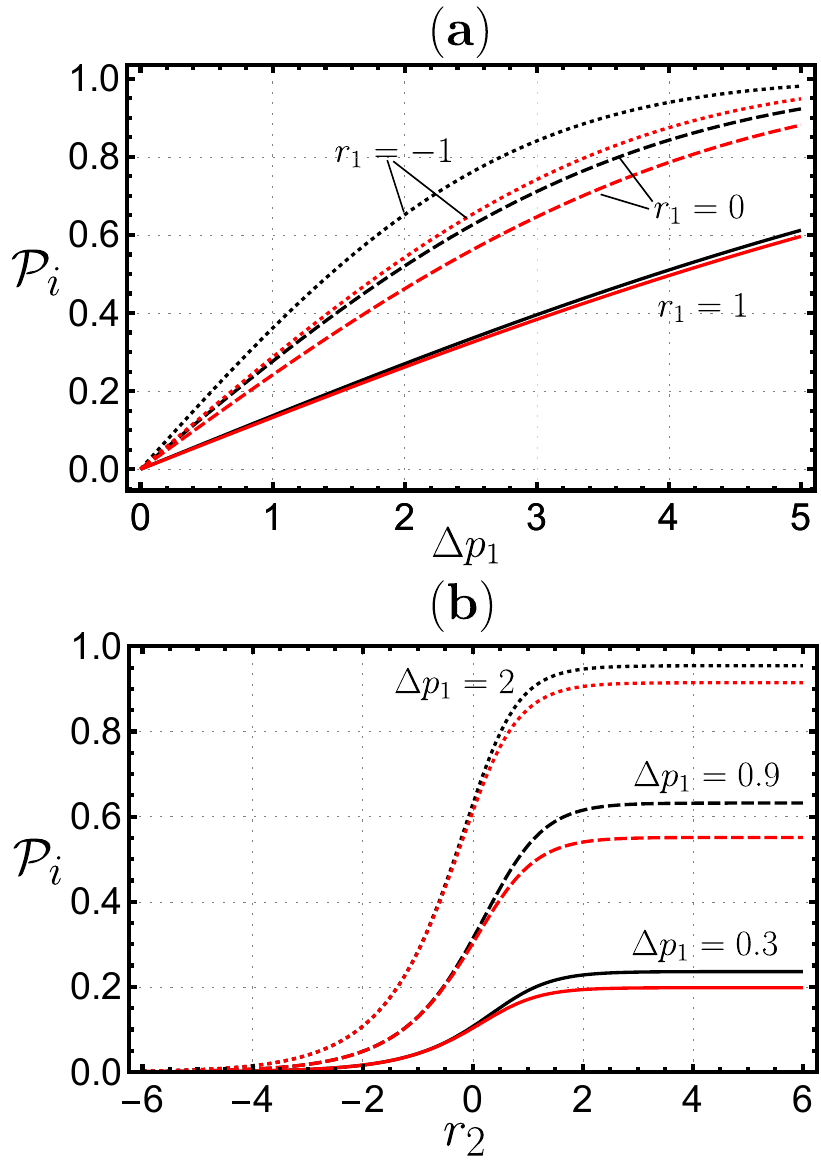}
\centering
\caption{Probability of teleportation of the squeezed-coherent state versus (a) the width of the selectivity interval $\Delta p_{1}$ ($V_{s}^2 = 1, \theta =\pi$) for different squeezing parameters $r_{1}$, and (b) the squeezing parameter of the squeezed vacuum state ($\delta_{q}^2 (r_{1}, \theta) = 0.25, V_{s}^2 =e^{2r_{2}}$) for various widths of the selectivity interval. The black curves represent the case where $\Delta p_{1}$ is centered on the central coordinate of the position of the squeezed-coherent state ($p_{1}'=-q_{0}$), while red curves represent the case where $\Delta p_{1}$ is not centered on the momentum coordinate allowing maximum teleportation; in particular for (a) we have $(p_{1}'+q_{0})=0.75$ and (b) $(p_{1}'+q_{0})=0.3$. Figures (a) and (b) show that the probability of teleportation increases when the selectivity interval of the measurement and the squeezing in position of both states building the cluster increases; besides, as we move away from the central value $p_{1}'=-q_{0}$, the probability of teleportation decreases.}\label{prob}
\end{figure}
where we recall that $X_{0}=q_{0} + p_{1}$ is the effective displacement, and $\sigma=\left\lbrace V_{s}^2/\left[1+ V_{s}^2\delta_{q}^{2}(r_{1}, \theta)\right] \right\rbrace^{1/2}$. The Eq. \eqref{eq:44.01} represents a quantitative measure of the degree of overlap between the probability distributions of both the squeezed vacuum state and the squeezed-coherent state under teleportation. Now, the $i$-esim detector of Fig. \ref{fig:00} will have associated a probability of detection—i.e., a likelihood to obtain the central measurement value $p_{1}'$ of the corresponding selectivity interval $\Delta p_{1}$—. By integrating Eq. \eqref{eq:44.01}, according to Eq. \eqref{eq:34}, we obtain such likelihood as
\begin{equation}
\mathcal{P}_{i}=\frac{1}{2}\left[ \text{erf}\left( \frac{\sigma Y_{+}}{\sqrt{2}}\right) -\text{erf}\left( \frac{\sigma Y_{-}}{\sqrt{2}}\right)\right], \label{eq:44.0001}
\end{equation}
where $\text{erf}\left(x\right)$ is the error function, $Y_{\pm}=\left(q_{0}+p_{1}'\right) \pm \Delta p_{1}/2$, and $p_{1}'$ is the central value of the selectivity interval $\Delta p_{1}$.
The Eq. \eqref{eq:44.0001} implies that the likelihood of getting measurement in the $i$-esim detector is determined by the width of its selectivity interval, its associated measurement result, and the squeezing properties of the states building the cluster. The measurement device with higher likelihood of detection will be the one whose associated outcome brings the higher overlap between the wave functions of the squeezed-coherent state and the squeezed vacuum state building the cluster. In our particular instance, we have this peak at $p_{1}'=-q_{0}$; by following this value, the expression of Eq. \eqref{eq:44.0001} reduces to
\begin{equation}
\mathcal{P}_{(p_{1}'=-q_{0})}=\text{erf}\left(\frac{\Delta p_{1} \sigma}{2\sqrt{2}} \right).\label{eq:44.02}
\end{equation}
\begin{figure}
\includegraphics[width=0.52\textwidth]{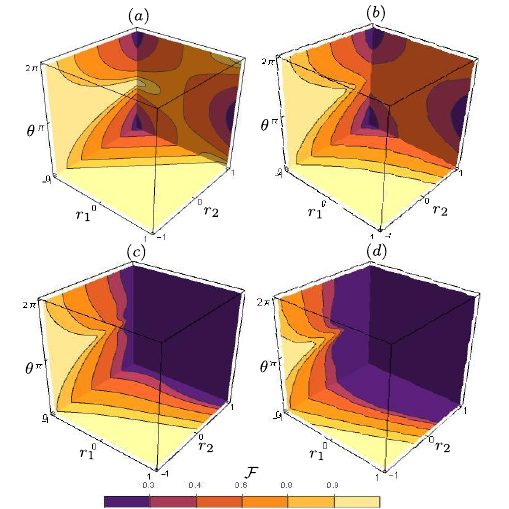}
\centering
\caption{Behavior of the fidelity of teleportation $\mathcal{F}$ of a squeezed-coherent state inside the three-dimensional region $\mathcal{R}=\left\lbrace (r_{1}, r_{2}, \theta) | -1\leq r_{1}\leq 1, -1\leq r_{2}\leq 1, 0\leq\theta \leq 2\pi\right\rbrace$ for a two-mode cluster state with finite squeezing. The effective displacements are (a) $X_{0}=0$, (b) $X_{0}=\pm 1$, (c) $X_{0}=\pm 1.75$, (d) $X_{0}=\pm 3.5$. The fidelity inside $\mathcal{R}$ diminishes as the effective displacement increases.}\label{fig:2}
\end{figure}
It is straightforward to verify that the increasing of $\sigma$ implies the increasing of the probabilities of Eqs. \eqref{eq:44.0001} and \eqref{eq:44.02}, which in turn implies the increment of the parameter $V_{s}^2$, meaning a higher squeezing in position for the squeezed vacuum state of the cluster. Besides, the increasing of $\sigma$ also implies the decreasing of the variance $\delta_{q}^{2}(r_{1}, \theta)$ of the squeezed-coherent state, where the rotation angle optimizing such decreasing is $\theta=\pi$, for the region $r_{1}<0$. 

In all the subsequent, we will utilize $V_{s}^2=e^{2r_{2}}$ for the squeezed vacuum state, being $r_{2} \in \mathbb{R}$ its squeezing parameter; therefore, such state will exhibits position (momentum) squeezing when $r_{2}>0$ ($r_{2}<0$); then, with these considerations, in Fig. \ref{prob} we plot the probability of Eq. \eqref{eq:44.02} versus (a) the width of the selectivity interval of the measurement and (b) the squeezing parameter of the squeezed vacuum state of the cluster. From that plots it is verified that the probability of teleportation increases when the selectivity interval of the measurement and the squeezing in position of both states building the cluster increases. Besides, as we move away from the central value $p_{1}'=-q_{0}$, the likelihood for teleportation attenuates.
\begin{figure}[]
\includegraphics[width=0.48\textwidth]{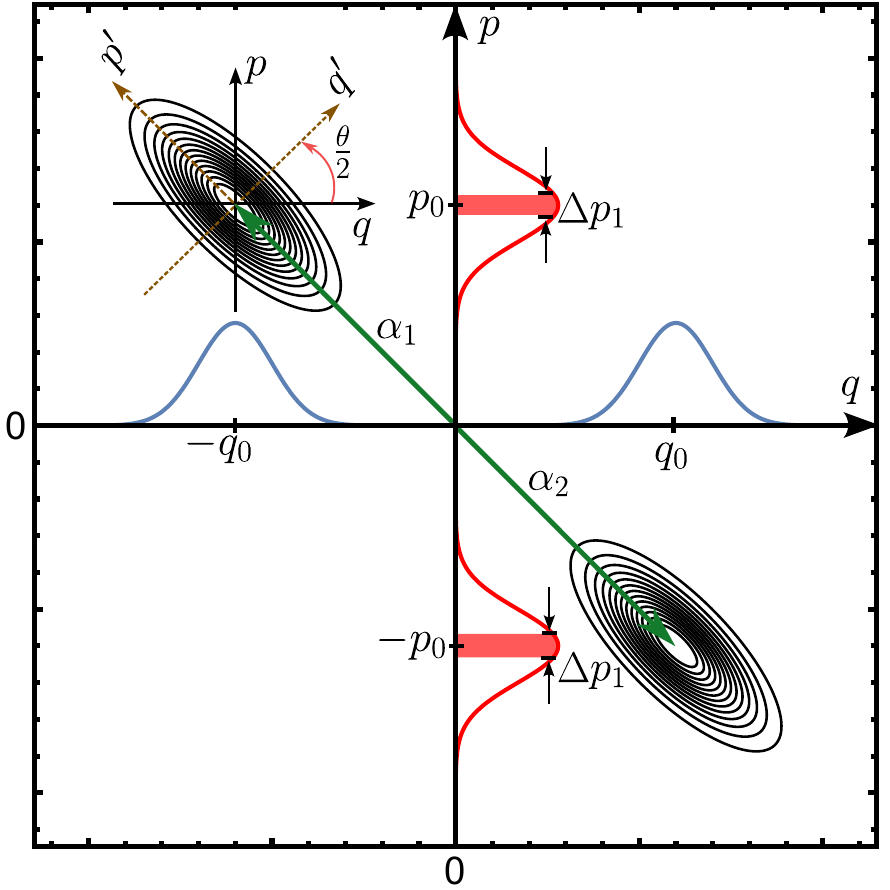}
\centering
\caption{Instance to enhance the fidelity of teleportation of the squeezed-coherent state through a two-mode cluster. We prepare the squeezed-coherent state in the second or fourth quadrant of the phase space with centers  $(-q_{0}, p_{0})$ and $(q_{0}, -p_{0})$ respectively, such that $\left|q_{0}\right|=\left|p_{0}\right|$; after building the cluster through the $\hat{C}_{z}$-gate, all the information of the momentum outcomes of the first mode is contained in the reduced density operator $\hat{\rho}_{1}$. Such a state has a Wigner function $W(q_{1},p_{1})$ (concentric ellipses) in phase space whose associated position (blue curves) and momentum (red curves) probability distributions are described by Gaussian centered in $\mp q_{0}$ and $\pm p_{0}$ respectively. The output that one single detector can register is determined by the central coordinate of its selectivity interval $\Delta p_{1}$, which encompasses a portion of the domain of the momentum probability distribution associated with $\hat{\rho_{1}}$ (red filling under the red curve); see also the scheme of Fig. \ref{fig:00}. Then, to get the maximal fidelity of teleportation, we must activate that detector whose selectivity interval is centered on $\pm p_{0}$; besides, we need to determine the measurement result with accuracy, which implies to take $\left|\Delta p_{1}\right|$ small enough.}\label{fig:3}
\end{figure} 
Notably, for the teleportation of the squeezed-coherent state, the detector corresponding to the maximum probability of equation \eqref{eq:44.02} is also associated with the maximum fidelity of teleportation. This will be discussed in detail in the following subsection.
\subsubsection{Fidelity of teleportation} \label{2.4.2}
The fidelity of individual teleportations in our approach is given by Eq. \eqref{eq:38.2}; then, we use the function $f_{G}(q)$ of Eq. \eqref{eq:31} and the probability distribution of Eq. \eqref{eq:44} to evaluate the integral of such expression; through this process, we obtain the fidelity of teleportation of the squeezed-coherent state in the $i$-esim detector as 
\begin{equation}
\mathcal{F}=\left[ P(p_{1})\right]^{-1}\left(\frac{2}{\pi V_{s}^2}\right)^{\frac{1}{2}} \delta^{2} e^{-\left(\delta X_{0}\right)^2}, \label{eq:45}
\end{equation}
where $P(p_{1})$ is the momentum probability distribution of Eq. \eqref{eq:44.01}; besides, $\delta =\left\lbrace V_{s}^2/ \left[2 + V_{s}^2 \delta_{q}^2(r_{1}, \theta)\right] \right\rbrace^{1/2}$. In Fig. \ref{fig:2} we show $\mathcal{F}$ inside the three-dimensional region  
\begin{equation}
\mathcal{R}=\left\lbrace (r_{1}, r_{2}, \theta) | -1\leq r_{1}\leq 1, -1\leq r_{2}\leq 1, 0\leq\theta \leq 2\pi\right\rbrace, \label{eq:47}
\end{equation}
for different values of the magnitude of $X_{0}$; from that figure, it is evident that the fidelity shows a diminishing trend as the magnitude of $X_{0}$ increases. This observation is mathematically substantiated by considering the limit: $\lim_{X_{0}\to \pm \infty} \mathcal{F} =0$, which is an expected result since—as can be verified from Eq. \eqref{eq:45}—the effective displacement comes into play through a Gaussian function centered in $X_{0}=0$, which attenuates the fidelity of teleportation as the magnitude of $X_{0}$ grows; this behavior indicates that the fidelity of teleportation for the squeezed-coherent state is influenced by the measurement outcome $p_{1}'$ just as is established in Ref. \cite{Gu2009}. Then, by considering the scheme of Fig. \ref{fig:00}, we can reduce the noise due to finite squeezing in the teleportation process by turning on only that detector whose corresponding measurement result maximizes the integral of Eq. \eqref{eq:38.2}. To illustrate this idea in the context of phase space, we board a straightforward instance. Let us consider that the squeezed-coherent state we want to teleport is prepared in either the second or fourth quadrant of phase space with amplitudes $\alpha_{1}=-q_{0} + i p_{0}$ and $\alpha_{2}=q_{0} - i p_{0}$ respectively, such that $\left|q_{0}\right|=\left|p_{0}\right|$; hence, the central coordinates of this state are $(-q_{0}, p_{0})$ and $(q_{0}, -p_{0})$. Subsequently, we apply the $\hat{C}_{z}$ gate to entangle this state with the squeezed vacuum state, obtaining the state of Eq. \eqref{eq:23}; then, the information about the possible momentum measurement results obtainable in the first mode is contained in the corresponding reduced density operator $\hat{\rho}_{1}=\text{Tr}_{2}\left[\hat{\rho}_{12} \right]$, where $\hat{\rho}_{12}=\hat{C}_{z}\left|\psi\right\rangle\left|0, V_{s}\right\rangle \left\langle V_{s}, 0\right| \left\langle \psi \right|\hat{C}_{z}^{\dagger}$ is the global density operator of the cluster. Notably, we can obtain the position and momentum statistics of $\hat{\rho}_{1}$ through the Wigner quasi-probability distribution \cite{Hillery1984}; that is, by taking into account a position basis $\left\lbrace \left| x \right\rangle \right\rbrace$ for the first mode of the cluster, and the definition for the Wigner function (units of $\hbar=2$) \cite{Case2008}: $W(q,p)=\left(4\pi \right)^{-1}\int du~ e^{iqu/2} \left\langle p + u/2 \right|\hat{\rho}\left|p-u/2 \right\rangle $, we obtain the 
\begin{figure*}[]
\includegraphics[width=0.9\textwidth]{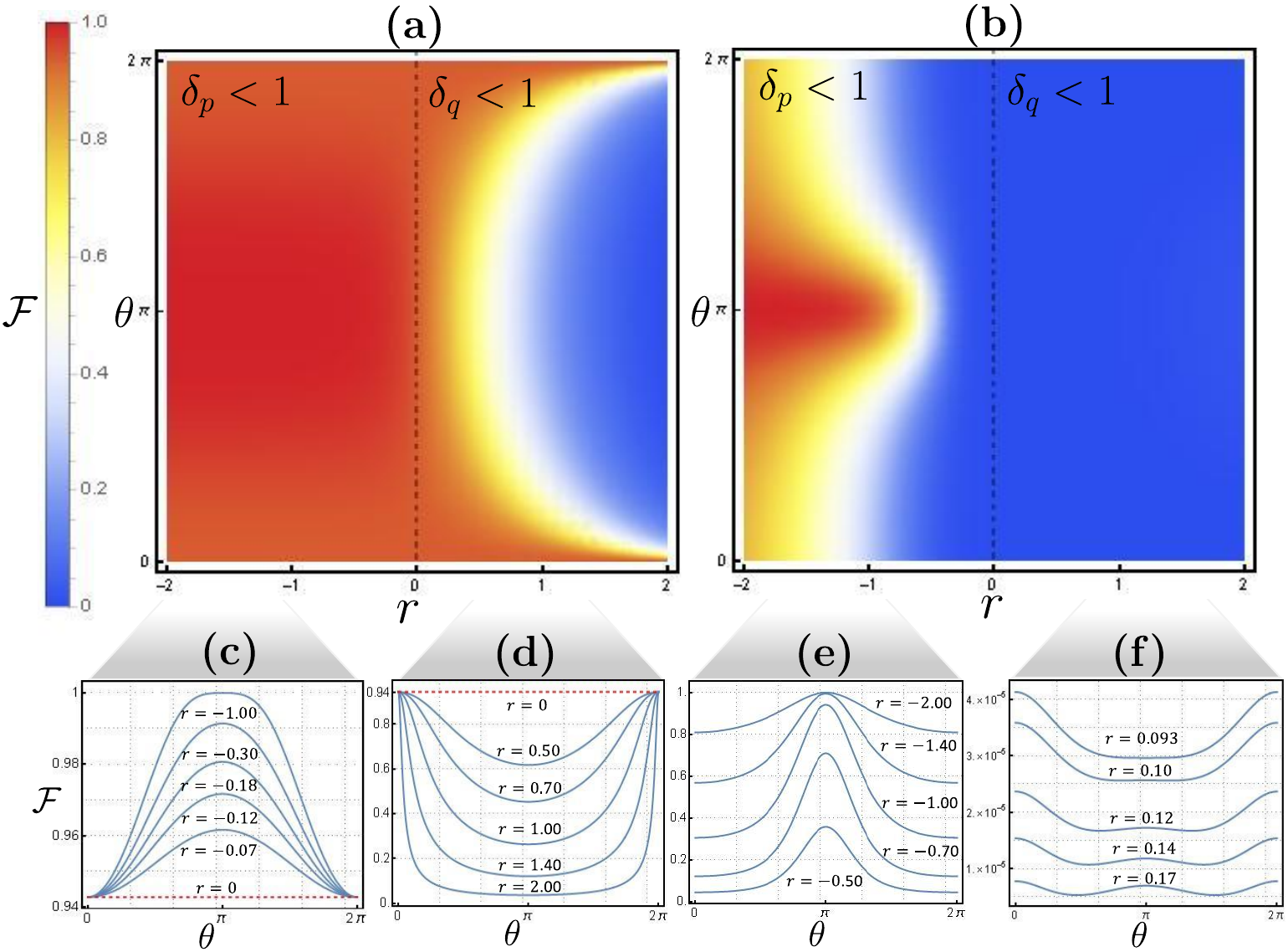}
\centering
\caption{Density plots (figures (a) and (b)) and curves (figures (c) - (f)) for the fidelity of teleportation $\mathcal{F}$ of a squeezed-coherent state through a two-mode CV cluster with finite squeezing. The squeezing parameters of the two states are $r_{1}=r_{2}=r$. The plots (a) and (b) describe the regimes of zero and non-zero effective displacements using $X_{0}=0$ and $X_{0}=\pm 10$ respectively. Each plot is divided into the regions $r>0$ and $r<0$ by the dotted line of $r=0$, pointing out the cases where the squeezed vacuum state of the cluster presents squeezing in the position and momentum probability distributions respectively; besides, this line represents the case where both states are coherent states. Figures (c)-(f) describe the behavior of the curves of fidelity as a function of the rotation angle $\theta$ for distinct $r$ parameters inside the regions $r>0$ and $r<0$. Then, as the effective displacement increases, the region of low fidelity (blue colored) expands towards the region of $r<0$, indicating that the squeezed vacuum state of the cluster requires progressively more squeezing in momentum to reach fidelities near to one.}\label{fig:4}
\end{figure*} 
joint probability distribution for position and momentum of $\hat{\rho_{1}}$ as
\begin{equation}
\begin{split}
W(q_{1},p_{1})=&\frac{1}{2\pi \hbar}\int dx du~e^{i q_{1} u/2}\left[f_{G}(x)\right]^2\\
& \times \psi(x - p_{1} - u/2)\psi^{\ast}(x - p_{1} + u/2);\label{eq:47.1}
\end{split}
\end{equation}
then, all information related to the potential measurement outputs is contained in the momentum probability distribution associated with the Wigner function of Eq. \eqref{eq:47.1}; that is, $P(p_{1})=\int dq_{1}~W(q_{1},p_{1})$; therefore, we must select the detector whose selectivity interval $\Delta p_{1}$ be centered on the momentum coordinate that offers a greater overlap between the functions $f_{G}(q)$ and $\rho(q)$; for our specific instance, we must center it on $\pm p_{0}$; see Fig. \ref{fig:3} for a sketch of this mechanism. Besides, it is worth noting that optimizing the fidelity of teleportation requires  determine the measurement outcome accurately, which implies taking the selectivity interval of the measurement to be small enough. Notably, according to Eq. \eqref{eq:34}, this process necessarily carries the decrement in the probability of teleportation (see Fig. \ref{fig:0}). Therefore, we must conclude that the likelihood and the fidelity of teleportation present a trade-off behavior regarding the width of the selectivity interval of the measurement.

On the other side, if we chose a non-optimal detector—i.e., the effective displacement $X_{0}$ is not near zero in Eq. \eqref{eq:45}—, we will need to increase the momentum squeezing of the squeezed vacuum state to reach teleportation with high fidelity. This fact aligns with the concept of an ideal cluster wherein zero momentum eigenstates are utilized for achieving flawless teleportation \cite{Menicucci2006}. To graphically visualize such an argument, we investigate the scenario where the squeezed-coherent state and the squeezed vacuum state building the cluster share identical squeezing parameters; that is, $r_{1}=r_{2}=r$. This election allows us to visualize the influence of the effective displacement in the fidelity of teleportation while maintaining some degree of generality. Then, in Fig. \ref{fig:4}, we show the density plots for the fidelity of teleportation of the squeezed-coherent state for the cases of $X_{0}=0$ and $X_{0}= \pm 10$. In both scenarios, the higher fidelities of teleportation happen when the squeezed vacuum state presents squeezing in momentum ($r<0$); however, it must be noted that as the effective displacement increases, we will require greater squeezing in momentum to achieve high fidelity of teleportation. Remarkably, in the regime of $X_{0}=0$, the cluster consistently yields a fidelity of $2\sqrt{2}/3$ when both input modes are coherent states ($r=0$) (the red dotted lines in (c) and (d) of Fig. \ref{fig:4}); furthermore, this fidelity is also obtained when both the squeezed-coherent state and the squeezed vacuum state match in the squeezing orientation, that is for $\theta=0$ and $\theta=2\pi$. For our particular instance, the higher fidelities happen at the regime of $X_{0} \approx 0$; however, this not be necessarily true in general; that is, according to Eq. \eqref{eq:38.2}, the maximum fidelities will arise when the Gaussian wave function of the squeezed vacuum state has significant overlap with the probability distribution of the teleported state;  then, the experimentalist should take the more convenient preparation of the state to teleport to select that device that allows enhancing the quality of teleportation.

To conclude the Section, we emphasize that our proposed scheme is especially significant in reducing noise due to finite squeezing in the fundamental block of one-way quantum computation as proposed in Ref. \cite{Menicucci2006}.  The significance arises mainly from the potential experimental limitations associated with generating high levels of squeezing, where the achievable peak values are typically around 15 dB \cite{Vahlbruch2016}.
\section{Sequential teleportations} \label{Sec:3}
Quantum communication between two points of a large channel can be achieved more efficiently if the channel is divided into several segments \cite{Briegel1998}. This mechanism can lessen the degrading effects that affect the information propagation through realistic noisy channels; for instance, we can preserve the entanglement between the resources in each link of the chain, which results in a higher quality of transmitted information even despite the cumulative error due to the finite squeezing of the sources. In the following, we utilize the concept of insufficiently selective measurements to analyze the scenario where $n$ circuits like the one in Fig. \ref{fig:1} are concatenated together. The difference between our scheme and previous references \cite{Menicucci2006, Gu2009, Menicucci2010} is that we consider between each step of the chain the corrections for successful teleportation employed in the ideal case of a cluster with infinite squeezing \cite{Weedbrook2012}.  
\subsection{Probability of teleportation} \label{probability}
The Eq. \eqref{eq:34} shows that the probability of obtaining teleportation within the cluster of Fig. \ref{fig:1} is contingent upon three factors: (i) the selectivity interval of the insufficiently selective measurement, (ii) the degree of squeezing of the squeezed vacuum state of the cluster, and (iii) the convolution between the position probability distributions of the squeezed vacuum state and that of the teleported state. Notably, for a chain of $n$ consecutive circuits like the one in Fig. \ref{fig:1}, the input state in each cluster originates from the output of the previous one; consequently, all the prior factors have a direct impact on the probability of teleportation in the immediate subsequent cluster; in other words, the likelihood of teleportation in a determined cluster of the chain is dependent on the teleportation process in the previous clusters. To elucidate this, we calculate the probability of teleportation in the $n$th cluster of a linear chain of sequential teleportations.

To start, we define the input state entering the $n$th cluster; for this, we consider the quasi-selective approximation in each teleportation process (see subsection \ref{fidelof}); besides, between each link of the chain we implement the corrections $\hat{F}^{\dagger} \hat{X}^{\dagger}$ as we explain in sùbsection \ref{fidelof}; with this considerations, the input state of each cluster is
\begin{equation}
\begin{split}
\left| \psi_{\text{in}}\right\rangle=&N^{-1}\hat{\mathcal{G}}_{(n-1)} \cdots \hat{\mathcal{G}}_{(0)}\left| \psi\right\rangle, \label{eq:48}
\end{split}
\end{equation}
where the normalization constant is $N=\sqrt{\left\langle \psi \right|\hat{\mathcal{G}}_{(0)}^{\dagger} \cdots \hat{\mathcal{G}}_{(n-1)}^{\dagger} \hat{\mathcal{G}}_{(n-1)} \cdots \hat{\mathcal{G}}_{(0)}\left| \psi \right\rangle}$ \footnote{It can be verified that this normalization constant is equivalent to taking the normalization of each output state in the second line of each of the $n-1$ clusters.}. Besides,  $\hat{\mathcal{G}}_{(i)}=\hat{F}^{\dagger}\hat{X}^{\dagger}\left(p_{1}^{(i)} \right)\hat{\mathcal{M}}_{i}\hat{X}\left(p_{1}^{(i)} \right)\hat{F}$ (for $i\neq 0$) is the operator that summarizes the set of continuous-variable quantum gates involved through the teleportation process inside the $i$-esim cluster; moreover, we define $\hat{\mathcal{G}}_{(0)}=\mathbb{\hat{I}}$, such that the state entering to the first cluster is the normalized state $\left| \psi\right\rangle$, just as is described in the formalism of the section \ref{2.3.1}. Furthermore, $p_{1}^{(i)}$ is the momentum outcome obtained in the $i$-th cluster; therefore, this value belongs to the selectivity interval $\Delta p_{1}^{(i)}$. 

To establish the probability of teleportation, we must follow the same method of determining Eqs. \eqref{eq:34} and \eqref{eq:34.1}; therefore, we need simply substituting in such equations the position probability distribution associated with the state of Eq. \eqref{eq:48}. By harnessing the unitarity of the Fourier gate and using a momentum basis $\left\lbrace \left|-p\right>\right\rbrace$, we obtain by recurrence the position probability distribution associated with Eq. \eqref{eq:48}, it is
\begin{equation}
\varrho(q)=N^{-2}\prod_{i=0}^{n-1} \left[f_{G_{(i)}}\left(q +p_{1}^{(i)}\right)\right]^2 \rho(q),\label{eq:48.a}
\end{equation}
with the normalization constant
\begin{equation}
N=\sqrt{\int dq \prod_{i=0}^{n-1} \left[f_{G_{(i)}}\left(q + p_{1}^{(i)}\right)\right]^2\rho(q)},\label{eq:48.b}
\end{equation}
where $f_{G_{(i)}}(q)$ is the Gaussian wave function associated with the squeezed vacuum state of the $i$-esim cluster (see Eq. \eqref{eq:31}), and $\rho(q)$ is the density of probability in position space of the quantum state entering to the chain; besides, we define $f_{G_{(0)}}(X)\equiv 1$,  for $X  \in \mathbb{R}$; therefore $\varrho(q)=\rho(q)$ for the case of one single cluster. It is worth noting that the position probability density of Eq. \eqref{eq:48.a} includes a normalization constant as a difference of its equivalent of one single cluster; as is evident from Eq. \eqref{eq:48.b}, the behavior of such normalization constant depends on the overlap between the wave function summarizing the noise added by all previous the teleportation processes and that of the original state entering to the chain; then, as a consequence of this, we must expect a different behavior for the probability of teleportation in a sequence of clusters than the case of teleportation through a single one.

On the other hand, it is worth noting that the squared function $\left[f_{G}\left(q + p_{1}^{(i)}\right)\right]^2$ represents a Gaussian probability distribution function (PDF) centered around $p_{1}^{(i)}$. Notably, pairwise products of Gaussian PDFs result in another Gaussian PDF multiplied by a scale factor that has a Gaussian shape \cite{Bromiley2003}; then, this fact can be used recursively for the product of $\mathcal{N}$ Gaussian PDFs, which will give the general result of a Gaussian scale factor $S_{\mathcal{N}}$ multiplying a new Gaussian PDF $f_{G_{\mathcal{N}}}\left(q+\mathcal{P}_{1}^{(\mathcal{N})}\right)$ with variance $\delta_{q_{\mathcal{N}}}^2$ and center $\mathcal{P}_{1}^{(\mathcal{N})}$  \cite{Bromiley2003}; that is,
\begin{equation}
\prod_{i=0}^{\mathcal{N}} \left[f_{G_{(i)}}\left(q + p_{1}^{(i)}\right)\right]^2=S_{\mathcal{N}}\left[f_{G_{\mathcal{N}}}\left(q + \mathcal{P}_{1}^{(\mathcal{N})}\right)\right]^2,\label{eq:48.c}
\end{equation}
where the resulting Gaussian:
\begin{equation}
f_{G_{\mathcal{N}}}\left(q+\mathcal{P}_{1}^{(\mathcal{N})}\right)=\left( \frac{1}{2\pi \delta_{q_{\mathcal{N}}}^2}\right)^{\frac{1}{4}}\exp\left[-\frac{\left(q+\mathcal{P}_{1}^{(\mathcal{N})}\right)^2}{4 \delta_{q_{\mathcal{N}}}^2} \right]\label{eq:53}
\end{equation}
has variance and center respectively according to
\begin{equation}
\delta_{q_{\mathcal{N}}}^2= \left(\sum_{i=1}^{\mathcal{N}}\frac{1}{\delta_{q_{i}}^{2}}\right)^{-1}, \label{eq:54}
\end{equation}
\begin{equation}
\mathcal{P}_{1}^{(\mathcal{N})}=\left(\sum_{i=1}^{\mathcal{N}}\frac{p_{1}^{(i)}}{\delta_{q_{i}}^2}\right)\delta_{q_{\mathcal{N}}}^2;\label{eq:55}
\end{equation}
besides, the scale factor $S_{\mathcal{N}}$ of Eq. \eqref{eq:48.c} is given by 
\begin{equation}
\begin{split}
S_{\mathcal{N}}=&\frac{1}{(2\pi)^{(\mathcal{N}-1)/2}}\sqrt{\frac{\delta_{q_{\mathcal{N}}}^2}{\prod_{i=1}^{\mathcal{N}}\delta_{q_{i}}^2}}\\
&\times \exp\left\lbrace -\frac{1}{2}\left[\sum_{i=1}^{\mathcal{N}}\frac{\left(p_{1}^{(i)}\right)^2}{\delta_{q_{i}}^2} - \frac{\left(\mathcal{P}_{1}^{(\mathcal{N})}\right)^2}{\delta_{q_{\mathcal{N}}}^2}\right] \right\rbrace. \label{eq:56}
\end{split}
\end{equation}
The terms $\delta_{q_{i}}^2$ and $p_{1}^{(i)}$ in Eqs. \eqref{eq:54} to \eqref{eq:56} represent the variance of the position probability distribution of the squeezed vacuum state and the momentum measurement outcome both associated with the $i$-th cluster respectively; these quantities contribute to the resulting Gaussian PDF of Eq. \eqref{eq:53} as well as the Gaussian scale factor of Eq. \eqref{eq:56}; consequently, the probability for obtaining teleportation inside a determined cluster of the chain, is influenced by both the squeezing of each squeezed vacuum state and the measurement outcomes—i.e., each of the previous selectivity intervals—in the previous clusters; thus, according to this mechanism, we can conclude that the probability of teleportation through a chain of consecutive clusters is an example of conditional probability \cite{Niestegge2008}.

By utilizing Eqs. \eqref{eq:48.c} to \eqref{eq:56} in Eqs. \eqref{eq:48.a} and \eqref{eq:48.b}, the position probability distribution simplifies to
\begin{equation}
\varrho(q)=N^{-2} \left[f_{G_{(n-1)}}\left(q +\mathcal{P}_{1}^{(n-1)}\right)\right]^2 \rho(q),\label{eq:48.d}
\end{equation}
where
\begin{equation}
N=\sqrt{\int dq \left[f_{G_{(n-1)}}\left(q + \mathcal{P}_{1}^{(n-1)}\right)\right]^2\rho(q)};\label{eq:48.e}
\end{equation}
hence, the probability of teleportation in the $n$th cluster of the chain is obtained by the replace $\rho(q)\longrightarrow \varrho(q)$ in Eqs. \eqref{eq:34} and \eqref{eq:34.1}.

\subsection{Fidelity of teleportation}
To determine the fidelity of teleportation of an arbitrary state $\left| \psi \right\rangle$ in the $n$th cluster of the chain, we follow the procedure outlined in Subsection \ref{fidelof}. We construct the density operator associated with the output (normalized) state of the second mode of the $n$th cluster; for this, we utilize the state of the Eq. \eqref{eq:48} and take the change $(n-1)\longrightarrow n$ to construct the output density operator $\hat{\rho}_{\text{out}}=N^{-2}\left(\hat{\mathcal{G}}_{(n)} \cdots \hat{\mathcal{G}}_{(0)} \right)\left| \psi \right\rangle \left\langle \psi \right|\left(\hat{\mathcal{G}}_{(0)} \cdots \hat{\mathcal{G}}_{(n)} \right)^{\dagger}$; then, by employing Eq. \eqref{eq:35} and leveraging the unitarity of the Fourier gate in the products $\left(\hat{\mathcal{G}}_{(n)}\cdots \hat{\mathcal{G}}_{(0)}\right)$, we obtain the fidelity for sequential teleportations as $\mathcal{F}_{\text{seq}}=N^{-2}\left[\left\langle \psi \right| \hat{F}^{\dagger}  \left(\prod_{i}^{n }\hat{D}_{i}\right) \hat{F} \left|\psi \right\rangle \right]^{2}$ being $\hat{D}_{i}=\left[\hat{X}^{\dagger}\left(p_{1}^{(i)} \right)\hat{\mathcal{M}}_{i}\hat{X}\left(p_{1}^{(i)}\right)\right]$, and $N$ the constant of normalization; besides, the subscript `seq' stands for `sequential', labelling the fidelity for sequential teleportations. By taking into account that $\hat{F}\left|\psi \right\rangle =\int dp ~\psi(p) \left|p \right\rangle$ and the definitions for $\hat{\mathcal{M}}_{i}$ and $\hat{X}\left(p_{1}^{(i)} \right)$, we obtain by recurrence the fidelity for sequential teleportations; the result is
\begin{equation}
\mathcal{F}_{\text{seq}}=N^{-2}\left(\int dq~ \prod_{i=1}^n f_{G_{(i)}}\left(q+p_{1}^{(i)}\right)\rho(q)\right)^2,\label{eq:57}
\end{equation} 
where the normalization constant is
\begin{equation}
N=\left(\int dq~\prod_{i=1}^n \left[f_{G_{(i)}}\left(q+p_{1}^{(i)}\right) \right]^2\rho(q) \right)^{\frac{1}{2}};\label{eq:58}
\end{equation}
besides, $f_{G_{(i)}}\left(q \right)$ represents the Gaussian wave function associated with the $i$-th squeezed vacuum state, and $p_{1}^{(i)}$ is the momentum measurement outcome obtained in the $i$-th cluster; besides, $\rho(q)=\left| \psi(q) \right|^2 $ represents the position probability distribution of the teleported state.

The Eqs \eqref{eq:57} and \eqref{eq:58} can be simplified by using the product of Gaussians of Eq. \eqref{eq:48.c}; besides, by considering Eq. \eqref{eq:56} in Eqs. \eqref{eq:57} and \eqref{eq:58}, we obtain that the quotient between the set of scale factors implied by Eq. \eqref{eq:57} cancels; then, with these considerations, the fidelity of teleportation of Eq. \eqref{eq:57} reduces to
\begin{equation}
\mathcal{F}_{\text{seq}}=(N)^{-2}\left[\int dq~f_{G_{(n)}}\left(q+\mathcal{P}_{1}^{(n)}\right)\rho(q)\right]^2,\label{eq:59}
\end{equation}
with
\begin{equation}
N=\left(\int dq~\left[f_{G_{(n)}}\left(q+\mathcal{P}_{1}^{(n)}\right)\right]^2\rho(q)\right)^{\frac{1}{2}}, \label{eq:60}
\end{equation}
being $f_{G_{(n)}}\left(q+\mathcal{P}_{1}^{(\mathcal{N})}\right)$ the resulting Gaussian of Eq. \eqref{eq:53} ($n=\mathcal{N}$) which has center $\mathcal{P}_{1}^{(\mathcal{N})}$ and variance $\delta_{q_{\mathcal{N}}}^{2}$ of Eqs.  \eqref{eq:54} and \eqref{eq:55} respectively.

By considering the change of variable $q'=q+\mathcal{P}_{1}^{(n)}$ in Eqs. \eqref{eq:59} and \eqref{eq:60}, we can conclude that the fidelity of teleportation through a series of $n$ concatenated clusters is equivalent to the fidelity of teleportation through a single cluster (see Eq. \eqref{eq:38.2}) whose squeezed vacuum state has associated a Gaussian wave function $f_{G_{(n)}}\left(q+\mathcal{P}_{1}^{(n)}\right)$ in the position space. This result suggests that all components within the chain—namely, the sources generating the $n$ squeezed vacuum states and the $\hat{C}_{z}$ gates—can be consolidated into a unified configuration comprising a single squeezed vacuum source and a single $\hat{C}_{z}$ gate. This finding aligns with the results of Ref. \cite{Menicucci2010}, where is shown that a linear sequence of CV cluster states is equivalent to one single squeezing source and one single quantum non-demolition gate utilized to build the cluster. 
Then, given our result, all conclusions regarding the fidelity of teleportation through a single cluster remain valid (see Sec. \ref{fidelof}). Specifically, the fidelity of Eq. \eqref{eq:59} is still a quotient between the squared solution of the non-homogeneous heat equation of Eq. \eqref{eq:39} and the solution of the conventional heat equation; besides, the average fidelity of the chain also will be proportional to the fidelity of the individual teleportations (see Eq. \eqref{eq:40.23}).

On the other hand, one must expect that as the number of clusters in the chain increases, the fidelity reduces due to the accumulated Gaussian noise from all teleportation processes; of course, we can lessen the effect of this noise by increasing the squeezing of each squeezed vacuum state of the chain, but in addition—as we explain in Subsec. \ref{squeezed}—we can resort to the selectivity intervals of the measurements to select those outputs that provide a higher overlap between  $f_{G_{n}}\left(q+\mathcal{P}_{1}^{(n)}\right)$ and $\rho(q)$—i.e., that furnish a greater fidelity of teleportation—In the following section, we will harness this idea to show the handling of both the probability and the fidelity of teleportation in the scheme of sequential clusters by considering again the instance of a squeezed-coherent state as the quantum state under teleportation.
\subsection{Teleportation of a squeezed-coherent state through succesive clusters}
\subsubsection{Probability of teleportation}
\begin{figure}[]
\includegraphics[width=0.48\textwidth]{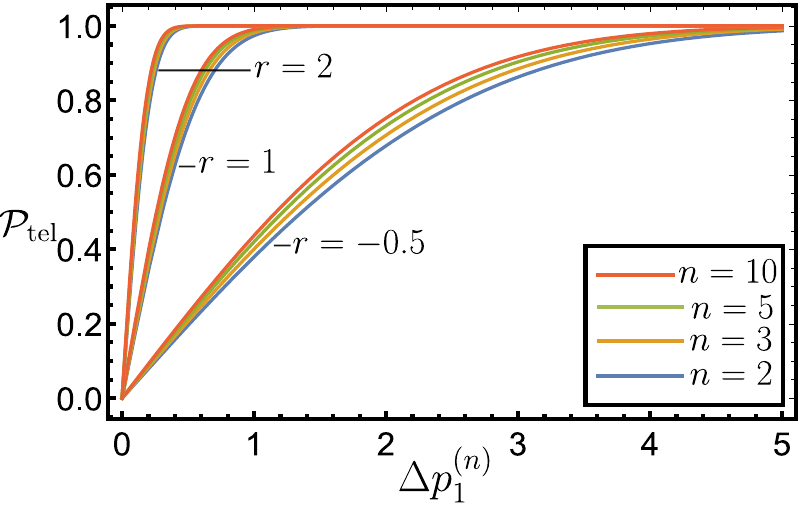}
\centering
\caption{Plots for the probability of teleportation of the squeezed-coherent state through a chain of $n=2, 3, 5, 10$ clusters for  $r=-0.5, 1, 2$ ($\delta_{q_{i}}^2=e^{-2 r}$, $\theta=0$, and $r_{1}=r_{2}=r$; therefore, $\delta_{q}^2(r_{1},\theta)=\delta_{q_{i}}^2$) versus the width of the selectivity interval of the $n$th cluster.  }\label{fig:13}
\end{figure}
To see the effect of sequential teleportations on the probability of teleportation, we examine a particular instance for the teleportation of the squeezed-coherent state of Eq. \eqref{eq:41} in the $n$th cluster of the chain. We assume a configuration where each squeezed vacuum state of the sequence has the same wave function: $f_{G_{(i)}}(q)=f_{G}(q)$ of Eq. \eqref{eq:31}, with a variance $\delta_{q}^{2}=e^{-2r_{2}}$; besides, we consider the case where the squeezed-coherent state under teleportation has the same squeezing that all squeezed vacuum states of the sequence (i.e., we take $\theta=0$ and $r_{1}=r_{2}=r$ in Eqs. \eqref{eq:42} and \eqref{eq:43}). Moreover, within each cluster, we activate the detector whose associated measurement result corresponds to the central position coordinate of the squeezed-coherent state undergoing teleportation; hence, under the quasi-selective approximation, each measurement result will be 
$p_{1}^{(i)} \approx q_{0}$, being $q_{0}$ the central coordinate of the state of Eq. \eqref{eq:41}. With the last assumptions, the parameters of Eqs. \eqref{eq:54} and \eqref{eq:55} reduce to
\begin{equation}
\delta_{q_{\mathcal{N}}}^2= \left(\mathcal{N} e^{2 r_{2}}\right)^{-1}, ~(\mathcal{N}\geq 1), \label{eq:61}
\end{equation}

\begin{figure*}
\includegraphics[width=0.99\textwidth]{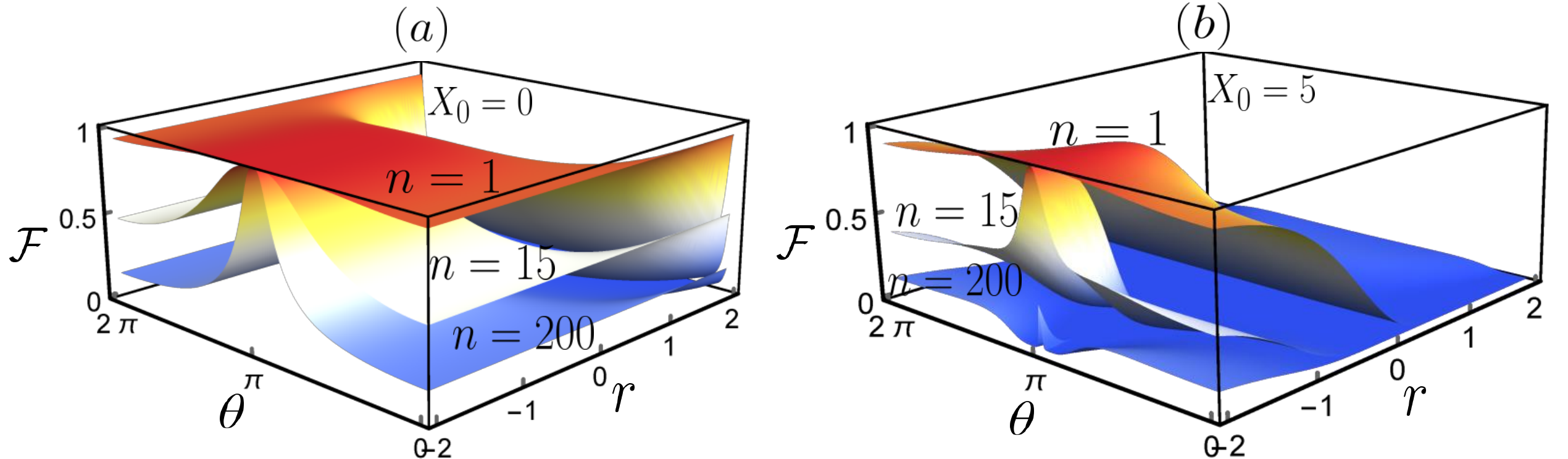}
\centering
\caption{Effect of attenuation in the fidelity of teleportation of a squeezed-coherent state through $n$ sequential clusters ($n=1$, $n=15$, and $n=200$) for the cases of effective displacements of (a) $X_{0}=0$ and (b) $X_{0}=5$. As the state is teleported through an increasing number of clusters, the fidelity decreases due to the Gaussian distortions induced in each step of the chain. The highest fidelities happen at the regime of zero effective displacement.}. \label{fig:5}
\end{figure*}  
\begin{equation}
\mathcal{P}_{1}^{(\mathcal{N})}=q_{0}.\label{eq:62}
\end{equation}
Now, we use Eqs. \eqref{eq:61} and \eqref{eq:62} in the resultant Gaussian of Eqs. \eqref{eq:48.d} and \eqref{eq:48.e}; then, we take the replace rule $\rho(q)\longrightarrow \varrho(q)$ in Eq. \eqref{eq:34.1}; moreover, we use the initial probability density of Eq. \eqref{eq:44}, where this time the displacement will be the measurement outcome in the $n$th cluster; that is, $p_{1}^{(n)}$. It is important to clarify that we are not using the definition of some effective displacement; however, we need to center the selectivity interval of the $n$th cluster of the chain such that the possible measurement results offer the maximum probability of teleportation. Then, with these considerations, we display in Fig. \ref{fig:13} the probability of teleportation of our particular scheme for $n=2, 3, 5, 10$;  for these cases, we carry out a numerical evaluation to find the optimal central value of the selectivity interval of the $n$th cluster; we find $p_{1}^{(n)}=0, 1/3, 3/5,$ and $ 4/5$ respectively. 
From Fig. \ref{fig:13}, we deduce that the probability of obtaining teleportation through the sequence increases with the increasing of the position squeezing of all states and the width of the selectivity interval of the $n$th cluster; this behavior is consistent with that of the teleportation process through a single cluster (see Section \eqref{2.4.1}). On the other hand, we have a higher probability of teleportation as the number of clusters of the chain is increasing; then, since the probability and fidelity of teleportation present a trade-off behavior (see Subsection \ref{2.4.2}), we would expect that the fidelity of teleportation decreases with a larger chain of clusters as well as the decreasing of the squeezing in momentum of the squeezed vacuum states of the chain; we will very this fact in the following subsection.
\subsubsection{Fidelity of teleportation}
To obtain the fidelity of teleportation of the squeezed-coherent state in successive clusters, it is worth noting that the expression of Eq. \eqref{eq:59} for the fidelity of teleportation through the chain maintains the same mathematical shape as in the case for a single cluster (see Eq. \eqref{eq:38.2} and consider the change of variable $q'=q+\mathcal{P}_{1}^{(n)}$ in Eqs. \eqref{eq:59} and \eqref{eq:60}); consequently, the fidelity of teleportation for a squeezed-coherent state through $n$ consecutive clusters is obtained simply by taking the replace rule: $\delta_{q}^2 \longrightarrow \delta_{q_{\mathcal{N}}}^2$ in Eqs. \eqref{eq:45} and \eqref{eq:44.01}, where $\delta_{q_{\mathcal{N}}}^2$ is the variance of Eq. \eqref{eq:54} and this time the effective displacement is defined in terms of the net displacement of the Eq. \eqref{eq:55}, that is $X_{0}=q_{0} + \mathcal{P}_{1}^{(\mathcal{N})}$.
\begin{figure}[]
\includegraphics[width=0.48\textwidth]{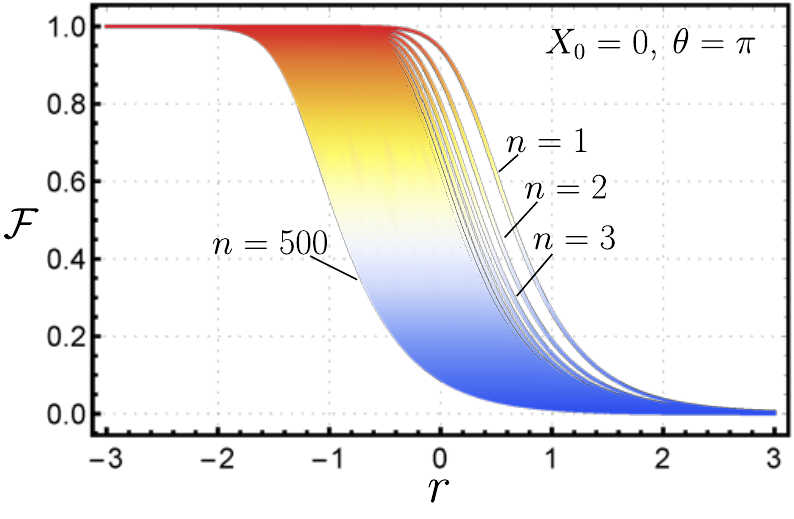}
\centering
\caption{Curves for the fidelity of teleportation ($X_{0}=0$, $\theta=\pi$) of a squeezed-coherent state for a chain of $n=1-500$ successive clusters. As the number of clusters in the chain increases, the peaks of the curves displace towards the region $r<0$, which means that a higher degree of squeezing in momentum for each squeezed vacuum state is required to reach high-quality teleportation ($\mathcal{F}\approx 1$).} \label{fig:6}
\end{figure}

To show the behavior of the fidelity of teleportation of the squeezed-coherent state in sequential clusters, we consider that all squeezed vacuum states of the chain have the same wave function: $f_{G_{(i)}}(q)=f_{G}(q)$, as is given by Eq. \eqref{eq:31}; besides, we assume that the squeezed-coherent state under teleportation has the same squeezing than all squeezed vacuum states of the chain; therefore we have $r_{1}=r_{2}=r$ and  $\delta_{q_{i}}^2=\delta_{q}^2(r,\theta)=e^{-2r}$.  With these considerations, in Fig. \ref{fig:5}, we show the plots describing the behavior of the fidelity of teleportation of a squeezed-coherent through $n=1, 15$, and $200$ successive clusters.  From these plots, we can conclude that as the squeezed-coherent state is teleported across a larger chain, the fidelity of teleportation will reduce; this effect comes from the accumulated Gaussian distortions that arise from each teleportation event in the chain; however, as we explain in Subsection \ref{squeezed}, such noise can be lessened through the handling of the effective displacement $X_{0}$; that is, it is possible to engineer a configuration where the squeezed-coherent state has an adequate localization in the phase space and then select appropriate selectivity intervals for the measurements in each cluster such that the net effective displacement converges to a value proximate to zero, where we will obtain the maximum fidelities of teleportation. 

On the other hand, as is verified from (a) and (b) of Fig. \ref{fig:5}, the maximums in the fidelity happen at the region of momentum squeezing ($r<0$): then, by considering the case of $X_{0}=0$, we carry out a numerical maximization to explore the rotation angle $\theta$ at which exist the peaks for the fidelity inside $r<0$ for $n=1-500$; in all cases, we find $\theta\approx\pi$. Notably, even with these appropriate parameters, we will need to increase more and more the momentum squeezing of the squeezed vacuum states of the chain to support high-quality teleportation over progressively longer chains; see Fig. \ref{fig:6}. Therefore, we verify again a trade-off behavior between the probability and the fidelity of teleportation, where the quality of the former is linked with the squeezing in position, while the second is with the squeezing in momentum.
\section{Conclusions} \label{Sec:4}
The present work uncovers key features concerned with the building block of one-way quantum computation with continuous variables; that is, the teleportation process within a two-mode continuous-variable cluster state. 

We take the fact that projective measurements in the continuous-variable regime can not be infinitely selective; there is a finite range of detection called the selectivity interval of the measurement apparatus over which all outcomes are essentially indistinguishable, giving all a unique central measurement result. By assuming that the whole measurement eigenspectrum is encompassed by a finite set of this kind of measurement devices, we derive the mathematical expression governing the probability of teleportation in each single detector of the configuration. We find that this likelihood constitutes the mirror image of the generalized Weierstrass transform of the position probability distribution associated with the system under teleportation; notably, this expression constitutes a valid solution of the heat equation.

On the other hand, by considering the quasi-selective approximation, we get the expression that describes the fidelity of teleportation for one detector in our approach; remarkably, we show that such expression constitutes a quotient between the squared solution of a non-homogeneous heat equation and that of the traditional heat equation. Besides, we find that both the fidelity and the probability of teleportation of a single cluster are in a trade-off relation regard: (i) the selectivity interval of the measurement, and (ii) the squeezing of the squeezed vacuum state. Particularly, by decreasing the width of the selectivity interval, we diminish the probability of teleportation, but the fidelity increases, and vice versa. Moreover, the likelihood of teleportation increases by increasing the squeezing in position of the squeezed vacuum state, while the fidelity rises with the growth of the squeezing in momentum.

Notably, both the probability and the fidelity of teleportation are directly dependent on the overlap (formally, the convolution) between the wave functions of the squeezed vacuum and the teleported state building the cluster; such overlap depends on the squeezing of the states, but also on the possible measurement outcomes attainable in the cluster; then, this fact situate the formalism of an insufficiently selective measurement apparatus as an adequate medium to handle the Gaussian noise affecting the quality of the teleportation in a two-mode continuous-variable cluster state since it allows to select those measurement results that increase the overlap mentioned earlier.

On the other hand, we consider a linear scheme of sequential teleportations with intermediate corrections. In particular, we obtain the exact expression governing the probability of teleportation in the sequence; it depends on the preparation of each squeezed vacuum state and each measurement result at each step of the chain. In addition, we determine that the fidelity of teleportation in this recursive model also depends on each of the Gaussian wave functions of the individual squeezed vacuum states building the chain; then, each teleportation process adds a Gaussian envelope that affects the quality of the whole fidelity. Besides, we find that the fidelity of teleportation maintains the same mathematical shape as in the case of teleportation through a single cluster; this means that our linear teleportation scheme is indistinct of a teleportation process involving one single cluster.

Furthermore, we find that both the probability and the fidelity of teleportation in a linear cluster also depend on the overlap between the resultant Gaussian function of the chain and the position probability distribution of the quantum state under teleportation; then, this fact also situates the insufficiently selective measurement formalism as a significant mechanism to select those outcomes that allow handling both the probability and the fidelity of teleportation in the scheme of successive teleportations. 

We have found a trade-off behavior between the probability and fidelity of teleportation in the linear sequence of clusters. Specifically, we observe that as the number of clusters in the chain increases, the fidelity decreases while the probability of successful teleportation increases, and vice versa. Besides, as in the case of one single cluster, the likelihood of teleportation increases with the growth of the position squeezing of the squeezed vacuum states of the chain, while the fidelity is higher with the increasing of the squeezing in momentum.

Finally, the present work provides the mathematical expressions to describe the quality of teleportation—probability and fidelity—in sequential teleportations, which serves as a medium to design and engineer efficient teleportation linear schemes through two-mode continuous-variable cluster states. Although our analysis is restricted to the Gaussian case, the extension to non-Gaussian resources constitutes a natural and relevant direction for future research, particularly in view of the distinctive features of these states compared to their Gaussian counterparts \cite{Sharma2025, Bose2021, Bose2021a}. 
\begin{acknowledgments}
J. A. Mendoza-Fierro thanks CONAHCYT for the
postdoctoral fellowship support under the application number 3762623.
\end{acknowledgments}
\bibliographystyle{apsrev4-2}
\bibliography{bibliography}
\end{document}